\documentclass[sigconf,10pt,nonacm,10pt]{acmart} 

\AtBeginDocument{%
  \providecommand\BibTeX{{%
    \normalfont B\kern-0.5em{\scshape i\kern-0.25em b}\kern-0.8em\TeX}}}

\setcopyright{acmcopyright}
\copyrightyear{2018}
\acmYear{2018}
\acmDOI{XXXXXXX.XXXXXXX}

\acmConference[Conference acronym 'XX]{Make sure to enter the correct
  conference title from your rights confirmation email}{June 03--05,
  2018}{Woodstock, NY}
\acmPrice{15.00}
\acmISBN{978-1-4503-XXXX-X/18/06}

\usepackage{xcolor}
\usepackage{color}
\usepackage{graphics}
\usepackage{graphicx}
\usepackage{url}
\usepackage{multirow}
\usepackage{xspace}
\usepackage{enumitem}
\usepackage{subcaption}
\usepackage{balance}
\usepackage{acmart-taps}
\usepackage[compact]{titlesec}

\usepackage{float}

\usepackage{xspace}

\newcommand{\remove}[1]{}

\newcommand{\pname}{Vivar\xspace}

\newif\ifflow
\flowtrue 

\newif\ifedit
\editfalse

\usepackage{calc}   
\usepackage{lipsum} 
\usepackage{multirow}
\usepackage{subcaption} 

\usepackage{ulem}

\usepackage[subtle]{savetrees}
\usepackage[skip=3pt]{caption}
\usepackage[subrefformat=parens]{subcaption}

\titleformat{\section}
  {\normalfont\Large\bfseries\MakeUppercase} 
  {\thesection}                              
  {1em}                                      
  {}                                         

\titlespacing*{\section}    
{0pt}{1.0ex plus 0.5ex minus 0.2ex}{0.5ex}
\titlespacing*{\subsection} 
{0pt}{0.7ex plus 0.3ex minus 0.2ex}{0.4ex}
\titlespacing*{\subsubsection} 
{0pt}{0.6ex plus 0.2ex minus 0.1ex}{1.0ex}

\definecolor{White}{rgb}{1, 1, 1}
\definecolor{Black}{rgb}{0, 0, 0}
\definecolor{Orange}{rgb}{0.902, 0.624, 0}
\definecolor{SkyBlue}{rgb}{0.337, 0.706, 0.914}
\definecolor{BluishGreen}{rgb}{0, 0.620, 0.451}
\definecolor{Yellow}{rgb}{0.941, 0.894, 0.451}
\definecolor{Blue}{rgb}{0, 0.447, 0.698}
\definecolor{Vermillion}{rgb}{0.835, 0.369, 0}
\definecolor{ReddishPurple}{rgb}{0.800, 0.475, 0.655}

\newcommand{\paragrapht}[1]{\vspace{1mm}\noindent{\textit{#1.}}\quad}

\newlist{compactitem}{itemize}{3} 
\setlist[compactitem,1]{label=\textbullet, leftmargin=*, topsep=0pt, itemsep=-1pt, parsep=1pt}
\setlist[compactitem,2]{label=--, leftmargin=*, topsep=0pt, itemsep=-1pt, parsep=1pt}
\setlist[compactitem,3]{label=*, leftmargin=*, topsep=0pt, itemsep=-1pt, parsep=1pt}

\newlist{compactenum}{enumerate}{3} 
\setlist[compactenum,1]{label=\arabic*., leftmargin=*, nolistsep} 
\setlist[compactenum,2]{label=\alph*., leftmargin=*, nolistsep} 
\setlist[compactenum,3]{label=\roman*., leftmargin=*, nolistsep} 

\setlist[itemize,enumerate]{left=0pt, itemsep=0pt, topsep=-2pt, parsep=0pt, partopsep=0pt}
\usepackage{quoting}
\quotingsetup{leftmargin=0pt, rightmargin=0pt, vskip=0pt}

\renewenvironment{quote}
  {\list{}{\leftmargin=0pt \rightmargin=0pt \topsep=0pt \partopsep=0pt \parsep=0pt \itemsep=0pt}%
   \item\relax\itshape} 
  {\endlist}

\makeatletter
\renewcommand\paragraph{\@startsection{paragraph}{4}{\z@}%
  {1.0ex \@plus.2ex \@minus.2ex}%
  {-.5em}%
  {\normalfont\normalsize\bfseries}%
  }
\renewcommand\paragraph[1]{\@startsection{paragraph}{4}{\z@}%
  {1.0ex \@plus.2ex \@minus.1ex}%
  {-.5em}%
  {\normalfont\normalsize\bfseries}{#1.}}
\makeatother

\ifflow
    \newcommand{\ppp}[1]{
    }
    \newcommand{\pp}[1]{
        \vskip 1ex\noindent
        \colorbox{yellow}{
            \parbox{\columnwidth - 2\fboxsep}{
                \textbf{Points:} #1
            }
        }
    }

\else
    \newcommand{\ppp}[1]{}    
    \newcommand{\pp}[1]{}    

\fi
\ifedit
    \newcommand{\tp}[1]{
        \vskip 1ex\noindent
        \colorbox{red}{
            \parbox{\columnwidth - 2\fboxsep}{
                \textbf{Issues:} #1
            }
        }
    }
    \newcommand{\preq}[1]{
        \vskip 1ex\noindent
        \colorbox{red}{
            \parbox{\columnwidth - 2\fboxsep}{
                \textbf{Pre-requisites:} #1
            }
        }
    }
    \newcommand{\yunqi}[1]{\textcolor{red}{YQ: #1}}
    \newcommand{\kaiyuan}[1]{\textcolor{red}{KY: #1}}
    \newcommand{\fred}[1]{\textcolor{red}{Fred: #1}}

    \definecolor{updatecolor}{RGB}{0,102,204} 
    \definecolor{deletecolor}{RGB}{255,0,0}   
    \definecolor{addcolor}{RGB}{0,153,0}      
    
    \newcommand{\delete}[1]{{\color{deletecolor}\sout{#1}}}
    \newcommand{\add}[1]{{\color{addcolor}{#1}}}
    \newcommand{\hc}[1]{\textcolor{red}{HC: #1}}
    \newcommand{\todo}[1]{\ClassWarning{NOT READY TO SUBMIT}{There is something left todo} \textcolor{blue}{[TODO: #1]}}

\else
    \newcommand{\tp}[1]{}    
    \newcommand{\preq}[1]{}    
    \newcommand{\yunqi}[1]{}
    \newcommand{\kaiyuan}[1]{}
    \newcommand{\fred}[1]{}

    \newcommand{\delete}[1]{}
    \newcommand{\add}[1]{#1}
    \newcommand{\hc}[1]{}

    \newcommand{\todo}[1]{\ClassWarning{NOT READY TO SUBMIT}{There is something left todo}}

\fi

\usepackage[subtle]{savetrees}

\begin{document}
\sloppy

\title[Visualizing the Invisible: A Generative AR System for Intuitive Multi-Modal Sensor Data Presentation]{Visualizing the Invisible: A Generative AR System \\for Intuitive Multi-Modal Sensor Data Presentation}





\author{Yunqi Guo\textsuperscript{*1}, Kaiyuan Hou\textsuperscript{*2}, Heming Fu\textsuperscript{1}, Hongkai Chen\textsuperscript{1}, Zhenyu Yan\textsuperscript{1},\\ Guoliang Xing\textsuperscript{†1}, Xiaofan Jiang\textsuperscript{†2}}
\affiliation{%
    \institution{\textsuperscript{1}The Chinese University of Hong Kong, \textsuperscript{2}Columbia University}
    \country{\{yqguo, hmfu, hkchen, zyyan, glxing\}@ie.cuhk.edu.hk, kh3119@columbia.edu, jiang@ee.columbia.edu}
}
\thanks{* Co-first authors, who contributed equally to this work.}
\thanks{† Joint corresponding authors.}


\renewcommand{\shortauthors}{LastName et al.}


\begin{abstract}
    Understanding sensor data can be difficult for non-experts because of the complexity and different semantic meanings of sensor modalities. This leads to a need for intuitive and effective methods to present sensor information. 
    However, creating intuitive sensor data visualizations presents three key challenges: the variability of sensor readings, gaps in domain comprehension, and the dynamic nature of sensor data. To address these issues, we propose \pname, a novel system that integrates multi-modal sensor data and presents 3D volumetric content for AR visualization. In particular, we introduce a cross-modal embedding approach that maps sensor data into a pre-trained visual embedding space through barycentric interpolation. This approach accurately reflects value changes in multi-modal sensor information, ensuring that sensor variations are properly shown in visualization outcomes. \pname also incorporates sensor-aware AR scene generation using foundation models and 3D Gaussian Splatting (3DGS) without requiring domain expertise. In addition, \pname leverages latent reuse and caching strategies to accelerate 2D and AR content generation, demonstrating $11\times$ latency reduction without compromising quality. A user study involving over 503 participants, including domain experts, demonstrates \pname's effectiveness in accuracy, consistency, and real-world applicability, paving the way for more intuitive sensor data visualization.
\end{abstract}

\maketitle

\section{Introduction} 
\label{sec:introduction}

Sensors have become indispensable in our increasingly interconnected world, capturing data beyond human perception and enabling deeper insights across fields such as environmental monitoring~\cite{Zhao_2023, aram2012environment}, healthcare~\cite{Dai_2023}, and industrial automation~\cite{Sprute_2023}. Sensors provide precise real-time data that are critical for informed decision making and process optimization. For example, environmental sensors help monitor climate change, assess pollution levels, and ensure safe living conditions~\cite{kang2020review,paradiso2003wearable}. In healthcare, sensors continuously track vital signs such as heart rate, oxygen levels, and body temperature, allowing for timely interventions and improved patient outcomes~\cite{helbig2021wearable, ko2010wireless}.

\begin{figure}
    \centering
    \includegraphics[width=1.0\linewidth]{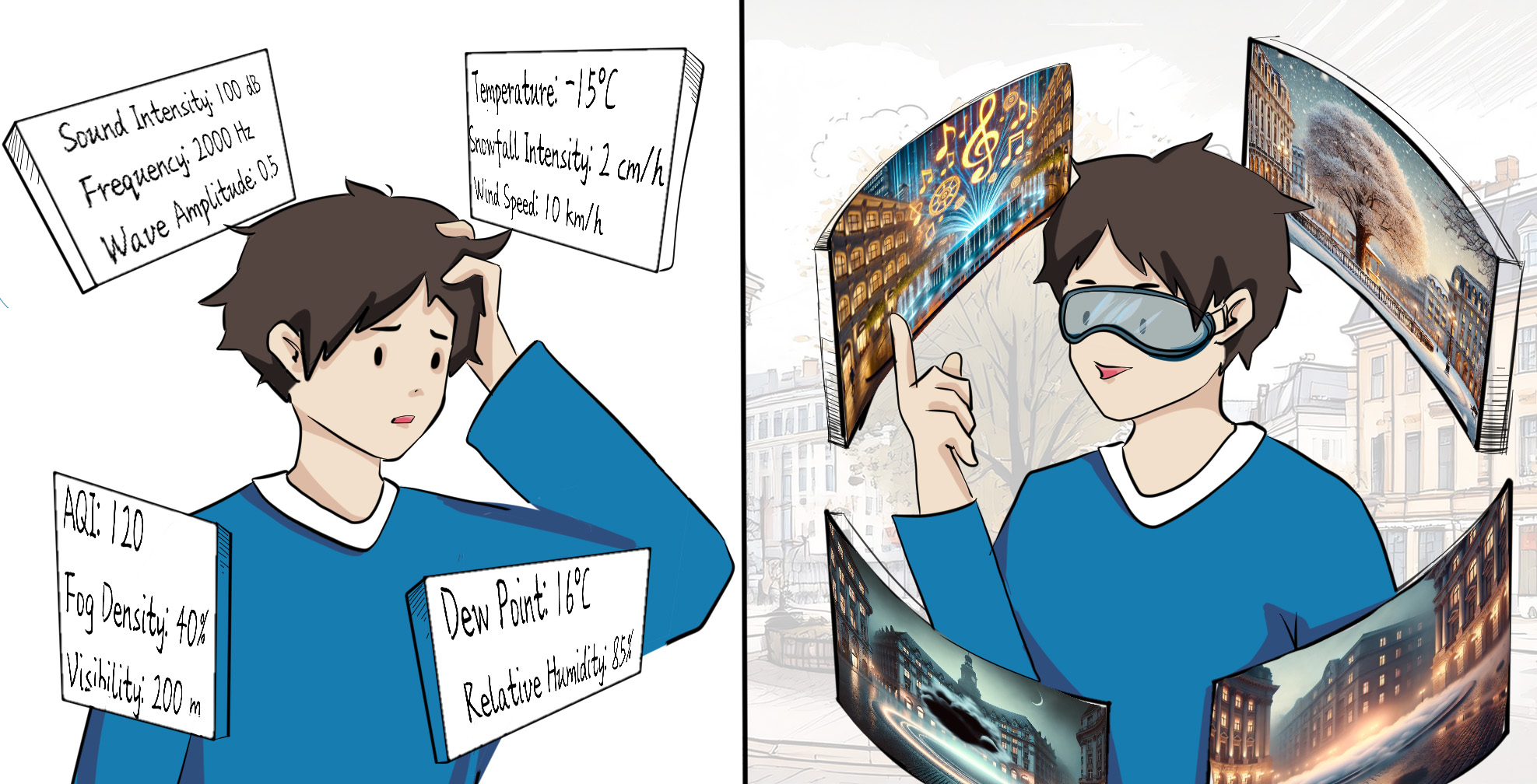}
    \caption{Left: without \pname, sensor data are abstract and hard to be interpreted by non-experts. Right: \pname generates AR scenes with intuitive and immersive sensor data presentation. }
    \vspace{-3mm}
    \label{fig:teaser}
\end{figure}

Although sensor data offers immense utility, its inherent complexity and abstract nature present significant challenges, particularly for non-experts seeking to interpret it. 
Raw data often contain intricate relationships, nuanced variations, and diverse modalities, making it difficult to derive meaningful insights at a glance. 
For example, an Air Quality Index (AQI) reading of 157, which indicates unhealthy air conditions, may be readily understood by environmental professionals but could confuse non-specialists who are less familiar with its implications for public health and outdoor activities.
To address this gap, intuitive and accessible visualization methods are essential, enabling transforming abstract data into interpretable forms that help better decision-making, situational awareness, and broader adoption of sensor-driven technologies.

Augmented Reality (AR) offers a particularly compelling medium for sensor information visualization to bridge this domain gap.
By overlaying digital content onto the physical world, AR provides an immersive and context-aware experience that enhances user interaction with data~\cite{ARapp1, ARapp2}. This capability allows users to intuitively engage with sensor information in their surroundings, turning complex, multimodal data into tangible, interactive experiences. For instance, AR can overlay environmental information such as air quality~\cite{ARairquality}, temperature~\cite{ARtemperature}, safety conditions~\cite{ARsafetyconditions}, or military training scenarios where the U.S. Air Force leverages multi-spectral operational mixed reality training devices to introduce new training efficiencies while overcoming previous technology limitations~\cite{Aechelon2023} as the representative manifestations onto physical spaces, offering a clear and accessible representation of these parameters without requiring specialized expertise. The immersive nature of AR effectively lowers barriers to data interpretation, benefiting both experts and non-experts, as shown in Figure~\ref{fig:teaser}.

Visualizing sensor information in AR to bridge the gap introduces three major challenges. 1)~\textit{Variability of Sensor Readings:} Sensor measurements often vary significantly, both individually and in combination. For example, temperature readings may range from -30°C to 40°C, representing vastly different conditions~\cite{temperatureDiverseRange}. Additionally, sensors frequently work together, measuring variables like air quality, temperature, and pollution levels simultaneously~\cite{SensorPollutionMonitoring}, making it challenging to coherently blend these interrelated data points into a unified scene. 2)~\textit{Domain Gap in Comprehension:} The abstract nature of sensor data creates a significant gap in human understanding, requiring tailored visual representations to effectively communicate key features. \add{Bridging this gap often involves long-term study and expert interpretation.} 3)~\textit{Dynamic Nature of Sensor Data:} The ever-changing nature of sensor readings necessitates just-in-time visualizations that match the rate of data updates, \add{requiring systems capable of rapid interpretation and immediate visual updates.}

Existing sensor data visualization methods face substantial limitations. Traditional dashboards, while effective for domain experts, often prove inaccessible to non-technical users. Recent advances in AI-driven visualization, such as Artificial Intelligence Generated Content (AIGC) frameworks like Stable Diffusion~\cite{esser2024scaling-sd3}, struggle to seamlessly integrate sensor data as inputs. Similarly, three-dimensional generation methods~\cite{tang2023dreamgaussian, poole2022dreamfusion, zhou2025autooccautomaticopenendedsemantic} show promise but are hindered by high computational demands and limited capability for generating volumetric sensor presentation. These shortcomings highlight the need for a new solution that combines intuitive and immersive user experiences with efficient performance.


To address these challenges, we propose \pname, a novel system for intuitive and efficient sensor data visualization in AR. Central to \pname is its use of \textbf{barycentric interpolation}~\cite{grunbaum1967convex,angel2005interactive}, a technique inspired by shading principles, to map diverse sensor data into a shared visual embedding space. This interpolation ensures smooth transitions and accurate blending of sensor readings, enabling seamless visualization in both single-sensor and multi-sensor scenarios. The system features an automated, sensor-aware pipeline that generates AR scenes directly from sensor readings, bridging the gap between abstract data and intuitive representation. To manage the rapid dynamics of sensor data, \pname incorporates a latent-reuse mechanism, caching 2D and AR content for timely responsiveness and enhanced efficiency.
\add{We have implemented \pname as an end-to-end system, encompassing data ingestion, visualization generation, and AR rendering. We will open-source our implementation to contribute to the field and inspire broader applications.}

We summarize our main contributions as follows:

\begin{enumerate}
    \item \textbf{Cross-modal embedding with barycentric interpolation}: We introduce a novel approach that maps diverse sensor data and combinations into a unified visual embedding space, enabling smooth transitions and improved interpretability.
    \item \textbf{Sensor-aware AR scene generation}: 
    We develop a robust system that automatically generates AR representations from the input of sensor data using foundational models and 3D Gaussian Splatting (3DGS) without requiring domain expertise.
    \item \textbf{Just-in-time efficiency}: We create a fast generation framework with latent-reuse and caching strategies for 2D and AR content that speed the generation up to $11\times$ compared to existing models.
\end{enumerate}

Extensive validation with over 450 participants, including domain experts, demonstrates the practical effectiveness of \pname in delivering accuracy, usability, and consistency across real-world applications.

\section{Related Work}
\label{sec:relatedwork}
\paragraph{Sensor Data Interpretation and Visualization}
Presenting the sensor data information to the users has been studied in both industry and academia, with tools like SensorViz~\cite{SensorViz} and immersive technologies for environmental data~\cite{ComparisonOfSpatialVisualizationTechInAR, CO2LLAB} providing detailed representations of specific sensor types in industries, smart cities, and traffic systems. However, these tools often cater to domain experts 
and fail to benefit non-expert users who struggle with interpreting complex and abstract sensor data.

Systems like DeepSee~\cite{DeepSee}, Virtual Dream Reliving~\cite{VirtualDreamReliving}, and ImmersiveFlora~\cite{ImmersiveFlora} overlay raw sensor data onto physical environments, improving interpretability in fields like environmental monitoring and healthcare. Other approaches use robots, avatars, or narrative techniques~\cite{FaceVis, CyborgBotany, Sensordoll, AvatAR, UsingComics}, yet they often lack seamless transitions across data modalities.


As access to sensor data grows, addressing these challenges is increasingly urgent. Effective visualizations enhance comprehension~\cite{GuesstheData, DoYouSeeWhatISee, ReadingBetweenthePixels, rinaldi2024art2musbridgingvisualarts, VRGamesExperience}, with methods like metaphorical mappings or prompt-based teaching~\cite{MetaphoricalVisualization, PromptingforDiscovery} showing promise. However, existing models are limited in handling heterogeneous sensor modalities. A system capable of intuitive, contextually relevant representations is essential for empowering both experts and the general public.

\paragraph{Cross-Modality Embedding}

Cross-modality embeddings unify diverse data types, such as text, images, and sensor data, into a shared representation space, enabling consistent interpretation of multimodal inputs. Models like ImageBind~\cite{girdhar2023imagebind}, Meta-Transformer~\cite{zhang2023metatransformer}, and OneLLM~\cite{han2024onellm} integrate heterogeneous data effectively but often require separate encoder for each modality and large paired datasets, limiting scalability in real-world sensor applications.
Furthermore, bias in pre-trained models can introduce semantic inconsistencies~\cite{BeyondTexttoImage}, leading to misunderstanding and wrong representations of the numerical sensor data. 

\paragraph{Generative Models in AR}
Generative models such as Stable Diffusion \cite{rombach2022high-stablediffusion} and DreamFusion \cite{poole2022dreamfusion} have gained prominence for their ability to create synthetic content and visualize complex data. These models excel in generating realistic imagery and 3D structures, with applications in domains ranging from art to scientific visualization~\cite{ImaginationCultivationinChildArtEducation}. While they demonstrate significant potential in bridging abstract data and tangible representations, their integration with environments with dynamic sensor readings remains underexplored.
Attempts to incorporate generative models into AR systems focus on creating immersive and adaptable 3D content. For instance, methods like SharedNeRF \cite{SharedNeRF} enable volumetric reconstructions for AR applications, offering realistic representations but facing challenges with computational overhead and latency. Models like 3DGS~\cite{kerbl3Dgaussians} have emerged to address these issues, improving efficiency for real-time applications. However, adapting these systems for sensor data visualization presents unique difficulties, such as maintaining semantic coherence while dynamically representing multimodal inputs~\cite{VisualIMU3DReconstruction, 3dshape2vecset3dshaperepresentation, Text3dGS, magic3d}.

The use of generative models for volumetric sensor content is still in its early stages, with most existing techniques focusing on pre-rendered or static visualizations. Our work tackles these limitations by incorporating a novel 3D content generation pipeline, enabling dynamic and intuitive representations of multimodal sensor data while addressing the latency and scalability challenges of current approaches.

\section{Motivation Study}
\label{sec:motivation}



\paragraph{Challenges in Understanding Sensor Data}
To understand how well people interpret sensor data and the necessity for visual representations, we conducted a survey with 122 participants across various age groups and technical backgrounds. Participants were presented with a series of common sensor measurements -- including air quality indices (AQI),  temperature, humidity, and ultraviolet (UV) index -- and asked to interpret these values and their implications for everyday decisions. Additionally, they evaluated alternative representation methods that used visual objects to embody the same data. Following are some examples in the survey:

\begin{enumerate}
    \item \textbf{Air Quality Index:} Participants were shown the reading ``AQI: 157'' and asked if outdoor activities were safe. Later, the same condition was presented as a visual scene with visible smog and reduced visibility.
    
    \item \textbf{Environmental Comfort:} Participants were asked to interpret the meaning of ``Temperature: 23°C, Humidity: 85\%'' for human comfort. The same condition was later visualized as a room with visible moisture on surfaces.
    
    \item \textbf{UV Exposure:} Participants interpreted ``UV Index: 8'' and recommended protective measures. This was later represented visually as intense sunlight causing visible effects on exposed surfaces.
\end{enumerate}

\begin{figure}
    \centering
    \includegraphics[width=0.99\linewidth]{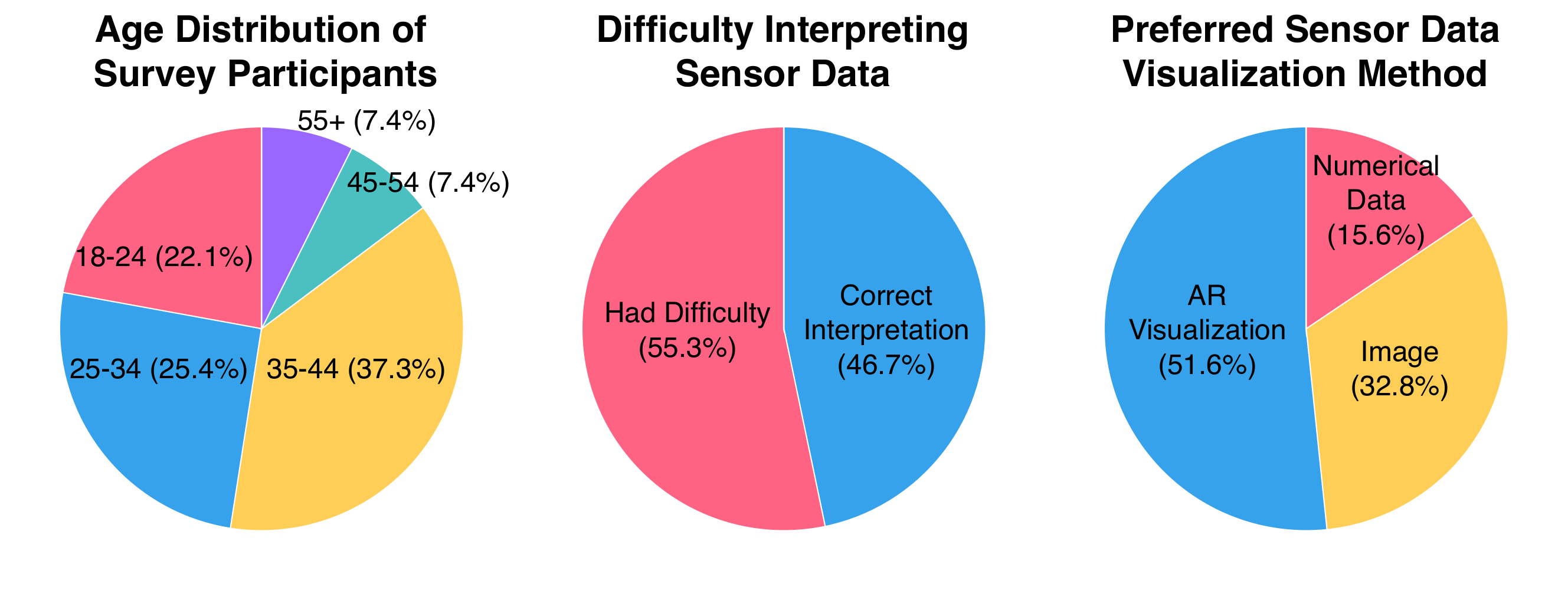}
    \caption{Survey About Sensor Data Interpretation}
    \label{fig:survey}
\end{figure}

As show in Figure~\ref{fig:survey}, this survey highlighted notable challenges for people in interpreting numerical sensor data: only 47\% of participants correctly identified safe versus unsafe AQI levels, while 38\% misinterpreted relative humidity readings. Additionally, 72\% of participants reported experiencing medium to high cognitive load when simultaneously integrating multiple sensor readings.
However, when presented with object-based visualizations of the same sensor data, comprehension accuracy improved by an average of 34\% across all sensor types. The most significant improvement (52\%) was observed among non-technical participants.
Participants were also asked about their preferences regarding different visualization approaches, specifically comparing AR-based visualization to traditional flat-screen methods. 
52\% of participants considered AR the most intuitive visualization method for experiencing sensor data, while 33\% preferred flat images, and 15\% preferred numerical data.
These results highlight both the difficulties people face when interpreting sensor data and the potential of intuitive visual representations -- especially in AR contexts -- to bridge this comprehension gap, particularly for users without domain expertise.


\paragraph{Foundation Model Limitations}
\label{sec:llm_limitation}
\begin{figure}[t]
    \parbox{\linewidth}{\centering \textbf{Context: The temperature is \textit{x} Celsius}} 
    \begin{subfigure}[t]{0.3\linewidth}
        \centering
        \includegraphics[width=\linewidth]{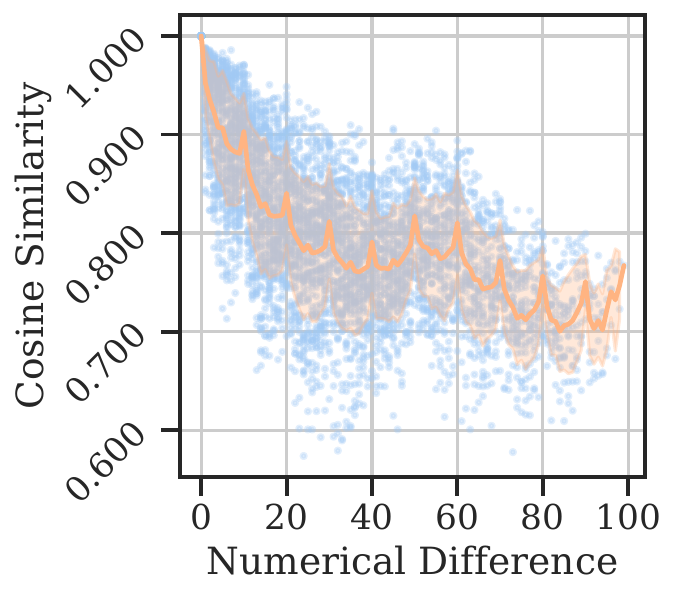}
        \Description{CLIP Temperature}
        \label{fig:clip_sim_temp}
    \end{subfigure}
    \hfill
    \begin{subfigure}[t]{0.3\linewidth}
        \centering
        \includegraphics[width=\linewidth]{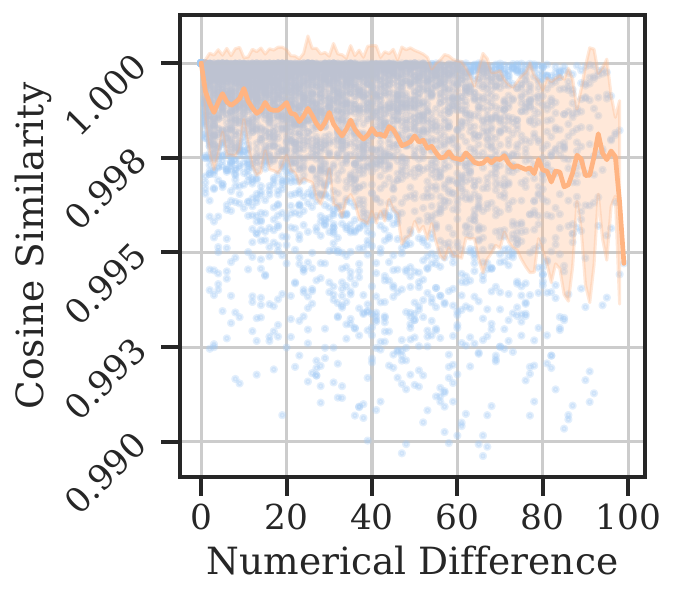}
        \Description{VisualBERT Temperature}
        \label{fig:vbert_sim_temp}
    \end{subfigure}
    \hfill
    \begin{subfigure}[t]{0.3\linewidth}
        \centering
        \includegraphics[width=\linewidth]{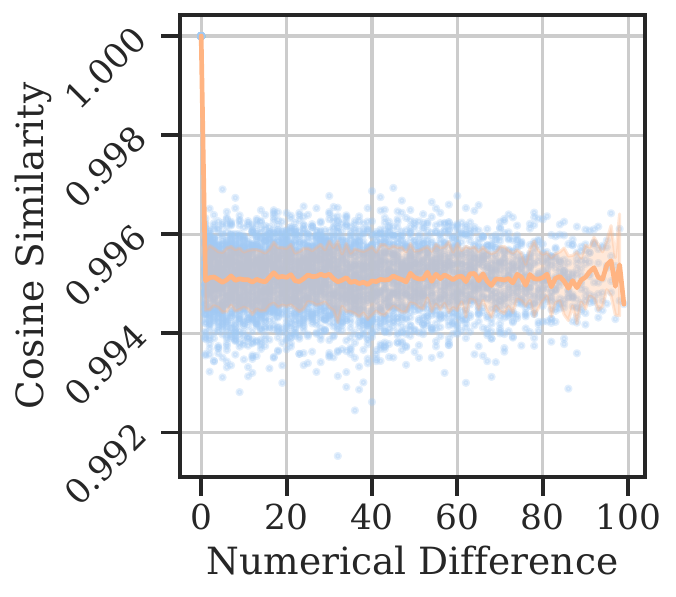}
        \Description{ALIGN Temperature}
        \label{fig:align_sim_temp}
    \end{subfigure}

    \parbox{\linewidth}{
        \centering
        \begin{minipage}[t]{0.32\linewidth}
            \centering \textbf{(a) CLIP}
        \end{minipage}
        \hfill
        \begin{minipage}[t]{0.32\linewidth}
            \centering \textbf{(b) VisualBERT}
        \end{minipage}
        \hfill
        \begin{minipage}[t]{0.32\linewidth}
            \centering \textbf{(c) ALIGN}
        \end{minipage}
    }

    \caption{Relationship between cosine similarity and numerical differences for three multimodal encoders. All three models demonstrate cognitive biases or limitations when embedding quantitative data.}
    \label{fig:diff_sim_combined}
\end{figure}

The previous survey demonstrates that visualization provides people with a more tangible way to understand abstract sensor data. However, creating effective visualizations presents a significant challenge: there is no comprehensive database that maps arbitrary sensor values to appropriate visual representations across diverse environments and contexts. This gap leads a generative approach that can dynamically create suitable visualizations tailored to specific sensor readings. This approach requires pretrained multimodal models that can convert numerical sensor data into compact and meaningful representations. These models enable the creation of unified embedding spaces where sensor readings can be seamlessly integrated with visual elements, facilitating the generation of contextually appropriate visualizations that enhance human comprehension of complex sensor data. 

\add{However, this approach requires models to be sensitive to semantic changes in numerical values. We conducted an experiment to assess how state-of-the-art multimodal encoders -- CLIP~\cite{radford2021learning}, VisualBERT~\cite{li2019visualbertsimpleperformantbaseline}, and ALIGN~\cite{jia2021scalingvisualvisionlanguagerepresentation} -- comprehend numbers in context. We encoded sentences like "The temperature is $x$ Celsius" into embedding spaces, and calculated the cosine similarity between embeddings relative to the numerical differences, as shown in Figure~\ref{fig:diff_sim_combined}. Each blue point represents:
$x = |v_1 - v_2|$, $y = \text{sim}(\mathcal{E}(\mathcal{C}(v_1)), \mathcal{E}(\mathcal{C}(v_2)))$
, where $\mathcal{C}(v)$ is the contextual sentence with value $v$, $\mathcal{E}$ is the encoder, and $\text{sim}$ is cosine similarity. The orange line shows the vertical average of these points.
Ideally, if the embeddings accurately reflect changes in numerical values, the relative relationships between numbers would be preserved. Specifically, larger numerical differences should correspond to lower cosine similarity in the embedding space. The results, shown in the figure, highlight several key issues: 
\begin{enumerate}[leftmargin=*]
\item[(a)] \textbf{Biases in embedding numerical readings}: While the overall trend decreases, the embeddings exhibit proximity to others at similar digits, reflecting periodic patterns that do not align with actual numerical relationships.
\item[(b)] \textbf{Inability to capture semantic changes}: The models fail to adequately reflect changes in sensor semantics, resulting in embeddings that are inconsistent with the underlying physical dynamics of the data. 
\end{enumerate}}

\add{Addressing these gaps requires a system that can effectively convert sensor data into intuitive and accurate AR visualizations through cross-modal embedding and sensor-aware generation methods.}



\section{Design Overview}
\label{sec:overview}

\add{Providing accurate and intuitive sensor information visualization presents three main challenges: 1) \textit{Multi-Modality:} Sensor readings often originate from diverse sources and formats, making seamless integration and interpretation challenging. 2) \textit{Context and Domain Differences:} The meaning of sensor data varies across different contexts, requiring tailored visual representations for accurate interpretation. 3) \textit{Temporal Dynamics:} Sensor data frequently change over time, necessitating visualizations that adapt to real-time changes.}

\begin{figure}[t]
    \centering
    \includegraphics[width=0.9\linewidth]{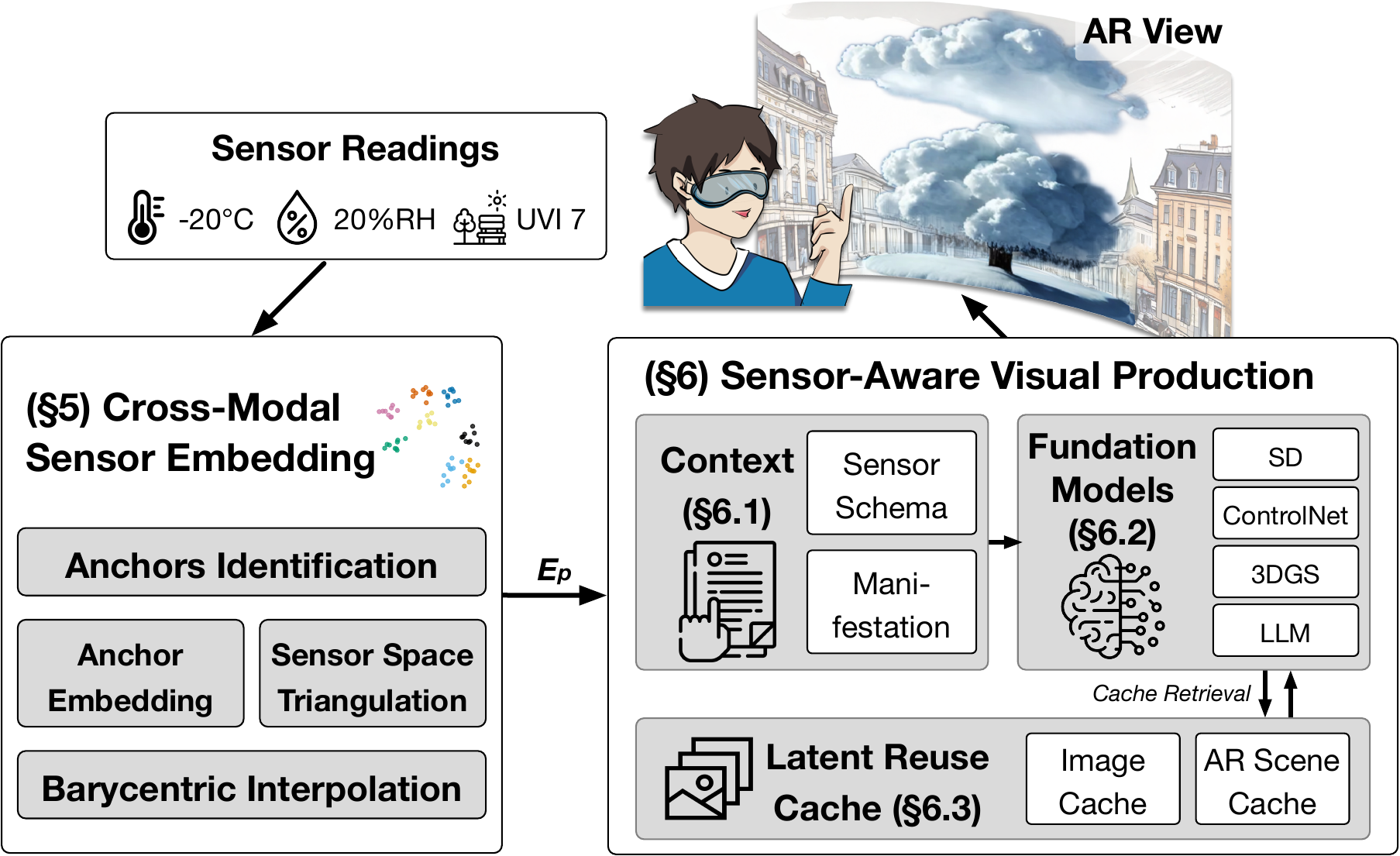}
    \caption{Overview of \pname Workflow.}
    \label{fig:workflow}
\end{figure}

To address these challenges, we present \pname, a system transforming abstract sensor data into intuitive and immersive AR experiences. 
As shown in Figure~\ref{fig:workflow}, the system operates in two key phases:

\begin{itemize}
    \item \textbf{Barycentric Interpolation for Cross-modal Sensor Embedding:} 
    We embed different types of sensor readings into a shared embedding space by first embedding representative sensor readings as anchors. 
    Our interpolation uses barycentric coordinates, inspired by shading principles, to calculate embeddings for new readings, ensuring smooth transitions while maintaining consistent representation across sensor types.
    
    \item \textbf{Sensor-Aware Visual Production:} Our sensor-aware visual production framework automatically integrates sensor context into visualizations. Using a customized volumetric 3D production model, we integrate spatial density of sensor information. Latent reuse caches at each production layer achieve response times 10\(\times\) faster than baseline 3D generation pipelines.
\end{itemize}




\section{Barycentric Interpolation for Cross-modal Sensor Embedding}
\label{sec:design1}

In this section, we introduce our approach for embedding multimodal sensor data into a shared space. 
Current embedding approaches have struggled with single modalities and the inability to handle numerical sensor data. 
To address these challenges, we proposed a barycentric interpolation method for the pre-trained visual-semantic embedding space.
First, the embedding space acts as a bridge between numerical sensor readings and their visual representations. Sensor data can be encoded into this space, while vectors in the space can be decoded into visual formats such as images or AR scenes.
Second, the embedding approach accommodates both individual sensor readings and their combinations. This representation must be accurate and consistent. Typically, similar sensor readings should produce similar visual representations, while readings with significant differences should yield distinct presentations. This principle also extends to combinations of readings, ensuring coherence and intuitive interpretation.

\subsection{CLIP as the Sensor Visual Space}
\label{sec:design-clip}
The embedding space must satisfy the following requirements: 
First, it should allow sensor information and context to be represented as vectors, with the capacity to capture the semantic information of the sensor readings. 
Second, these vectors must be transformable into visual representations. 
Given these requirements, a pre-trained language-image embedding model is an ideal candidate for our design.

Existing research demonstrates that Contrastive Language-Image Pre-Training~ (CLIP)~\cite{ramesh2022hierarchical-clip} effectively aligns text and image content by embedding both into a unified space. Originally designed for text and image data, this unified space enables CLIP to generate visual content based on embeddings~\cite{rombach2022high-stablediffusion, esser2024scaling-sd3, tang2023dreamgaussian}.

Despite its strengths, CLIP is primarily designed for text and image data, whereas sensor readings are numerical and present unique integration challenges as we have shown in Figure~\ref{fig:diff_sim_combined}. Mapping sensor information to visualization content requires a carefully designed function that translates raw data into meaningful visual representations within the shared ``visual space.'' The key objective is constructing a mapping function that accurately positions sensor data within this embedding space.

\begin{figure}
    \centering
    \includegraphics[width=0.9\linewidth]{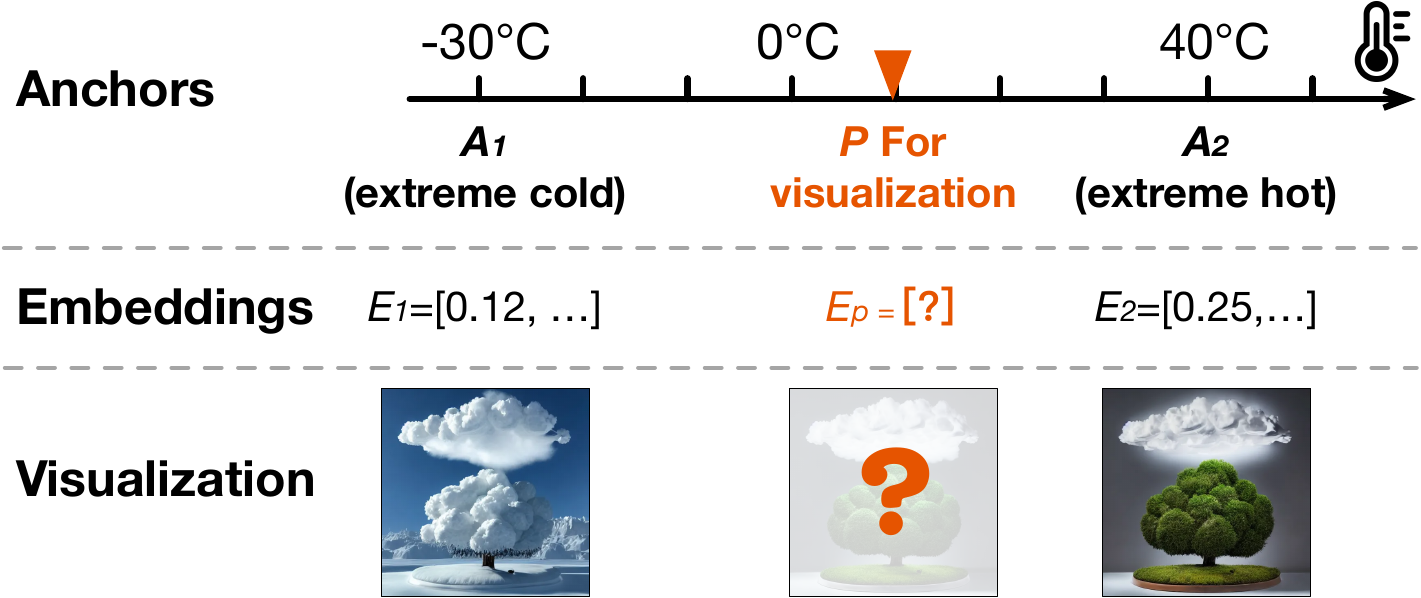}
    \caption{Anchors for embedding sensor readings.}
    \label{fig:anchor-illustration}
\end{figure}

\paragraph{Anchors and Interpolation for Embedding Calculation}
Previous studies~\cite{hou2024improving} have shown that interpolating between two embeddings enables image generation models, such as Stable Diffusion~\cite{huggingface2023stablediffusion}, to create smooth transformations from one generated image to another~\cite{huggingface2023stablediffusion, podell2023sdxl}. Inspired by this, we used an anchor-based interpolation method to generate embeddings for new sensor readings by interpolating between anchor embeddings.

Anchors are reference points that represent significant sensor readings in the visual space. They are chosen to capture key positions within the spectrum of sensor data, reflecting extreme or representative values of the sensor readings.

\begin{figure}[t]
    \centering
    \begin{subfigure}[t]{0.44\columnwidth}
        \centering
        \includegraphics[width=\linewidth]{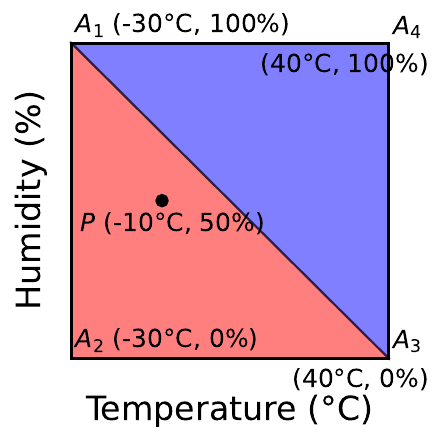}
        \caption{2 sensors with 4 anchors}
        \label{fig:temp_humidity_reading}
    \end{subfigure}%
    \hfill
    \begin{subfigure}[t]{0.44\columnwidth}
        \centering
        \includegraphics[width=\linewidth]{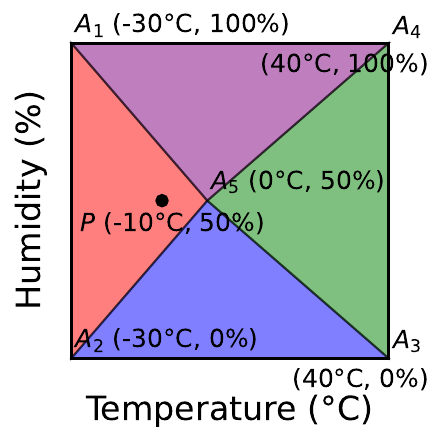}
        \caption{2 sensors with 5 anchors}
        \label{fig:temp_humidity_a5}
    \end{subfigure}%
    \caption{(a) Anchor points in the temperature-humidity space, with a new sensor reading $P$. (b) Introducing a new anchor point $A_5$ to the space.}
    \label{fig:temp_humidity_combined}
\end{figure}

The embeddings for these anchors are computed from the text embeddings of the sensor readings using a pre-trained CLIP text embedding model. Subsequently, these anchor embeddings are blended to generate embeddings for new sensor readings. As illustrated in Figure~\ref{fig:temp_humidity_reading}, for a single sensor reading, such as temperature, the embedding can be derived through linear interpolation between the corresponding anchors. For scenarios involving more than two anchors (e.g., with multiple representative points), the embedding is calculated using linear interpolation between the neighboring anchors (Figure~\ref{fig:temp_humidity_a5}).
Key challenges arise from the increasing number of sensors and anchors and the complexity of handling their combinations:

\begin{itemize}
    \item \textbf{Integration of Multiple Sensors:} Effectively combining readings from multiple sensors (e.g., temperature, humidity, light intensity) into a unified visual representation.
    \item \textbf{Ensuring Smooth Transitions:} Maintaining visual coherence and smooth transitions as sensor readings change, avoiding abrupt or inconsistent transformations.
\end{itemize}

\begin{figure}
    \centering
    \includegraphics[width=0.95\linewidth]{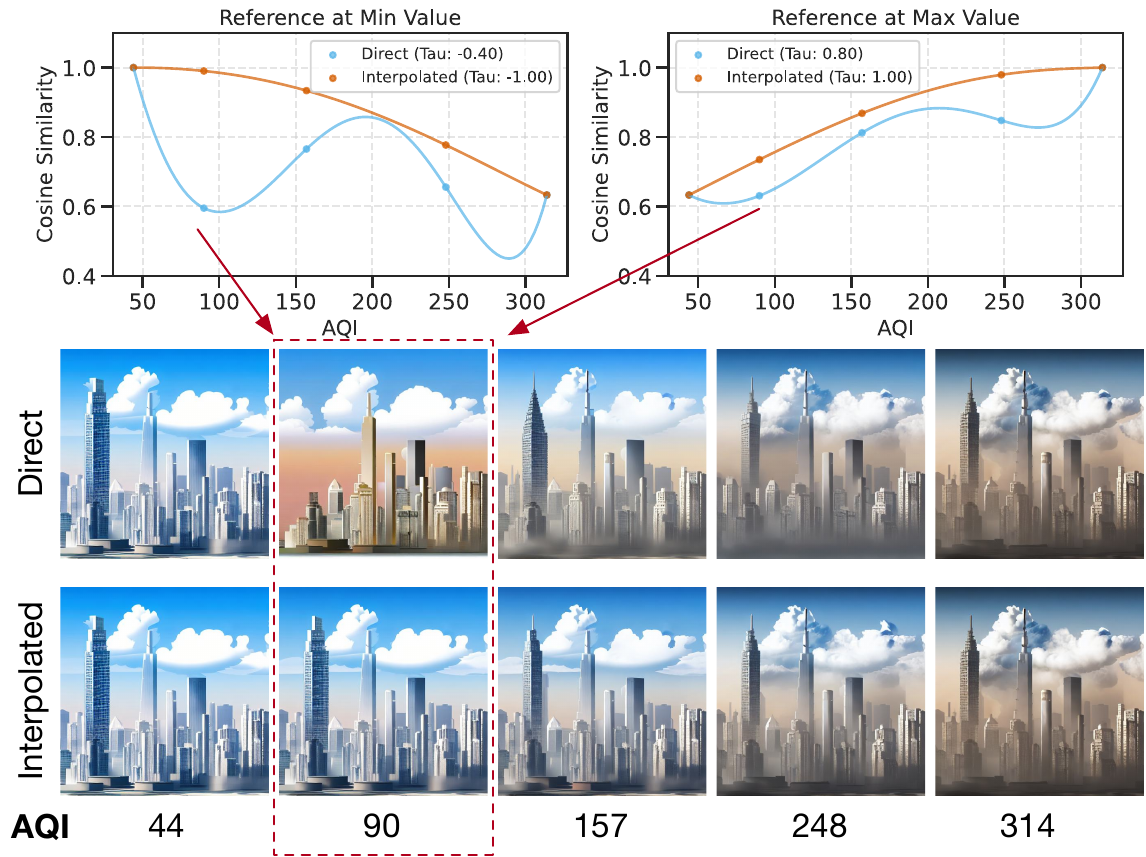}
    \caption{Embedding blinding enhances the monotonicity of similarity traces in both directions, with respect to the minimum and maximum boundaries.}
    \label{fig:aqi_skyline} 
\end{figure}

\paragraph{Showcase: Humidity Visualization}
We present a showcase example with humidity sensors to demonstrate the effectiveness of our method in visual space. 
We generate images using Stable Diffusion with embeddings computed by CLIP, based on the prompt template: \textit{"Urban skyline with buildings under \( x\) AQI}, where \( x\) can substitute with different values.
We compare the results of our embedding interpolation method (referred to as the `interpolated' set) with a baseline approach (referred to as the `direct' set), where embeddings were directly fed into the diffusion model. The comparative results are shown in Figure~\ref{fig:aqi_skyline}. For interpolation, we use embeddings generated at the minimum and maximum values of \( x \). 

As Air Quality Index (AQI) values increase, we expect to see a continuous progression of visual characteristics. These include a gradual transition in atmospheric coloration from clear blue to yellow-gray and finally to brown-red hues, as well as an increasing prominence of light source diffraction phenomena, resulting in the formation of halos or luminous effects through particulate matter. While this progression is evident in the interpolated set, we can clearly see this inconsistence in the direct set. Specifically, the image corresponding to AQI of 90 (the second image from the left) in the direct set fails to align with the expected progression.

To quantify these differences, we computed the cosine similarity of each embedding relative to the embeddings of the two extreme conditions (AQI of 44 and 314). This analysis yielded two sequences from the direct set, shown as blue lines in the top two plots in Figure~\ref{fig:aqi_skyline}. The corresponding sequences from the interpolated set are shown as orange lines in the same plots. 
To further evaluate the disruption in monotonic progression, we employed \textit{Kendall's Tau coefficient}, which measures the degree of monotonicity, with a value approaching 1 indicates a perfect monotonic increase, while a value approaching -1 indicates a perfect monotonic decrease. It is defined as 
\( \tau = \frac{2(P - Q)}{n(n - 1)} \), where \( P \) is the number of concordant pairs, \( Q \) is the number of discordant pairs, and \( n \) is the total number of pairs. 

The results indicate that the embedding interpolation method achieves perfect monotonicity, as reflected by Kendall's Tau values and the smooth progression of the orange curves. This contrasts with the inconsistencies observed in the direct set, as shown in the valley of the blue curve.

This example demonstrates that the embedding interpolation approach provides a more consistent and accurate representation of sensor data in the sensor visual space. By interpolation embeddings from anchor points, we achieve smooth transitions in the generated images corresponding to incremental changes in sensor readings. 
Next, we extend this design to cases with more than one sensor reading.

\subsection{Condition with Two Sensors}
Consider a scenario involving two sensors, such as temperature and humidity, where sensor readings are represented within two axes, and each axis corresponds to one sensor. Anchor points are defined to represent extreme combinations of these two dimensions, and their embeddings are calculated using pre-trained semantic models such as CLIP.


For instance, four anchor points represent the extreme combinations of temperature and humidity, as illustrated in Figure~\ref{fig:temp_humidity_reading}. These anchors are defined as follows: \( A_1 \) corresponds to low temperature and low humidity \((-30^\circ C, 100\%)\), \( A_2 \) represents low temperature and high humidity \((-30^\circ C, 0\%)\), \( A_3 \) denotes high temperature and low humidity \((40^\circ C, 0\%)\), and \( A_4 \) captures high temperature and high humidity \((40^\circ C, 100\%)\).

When a new reading, such as point \( P \) (e.g., -10°C, 50\% humidity), is received, an algorithm is needed to calculate \( P \)'s embedding relative to \( A_1 \) through \( A_4 \). In a 2D space, any three non-collinear anchor points can be used to interpolate \( P \). However, this flexibility introduces ambiguity, as the resulting combination for \( P \) may not be unique. Consequently, standard linear interpolation is unsuitable for determining \( P \)'s embedding.

To address this limitation, we propose a novel sensor embedding interpolation method based on barycentric coordinates.
This method divides the space of sensor readings into triangles and employs barycentric coordinates to calculate an intermediate embedding for \( P \) based on its surrounding anchor points, providing unambiguous and continuous embeddings for the sensor readings.

\paragraph{Barycentric Interpolation in 2D}

Barycentric coordinates provide a mathematical framework for representing a point within a simplex (e.g., a triangle in 2D) as a weighted combination of the simplex's vertices. The weights, or barycentric coordinates, uniquely define the point within the simplex and sum to 1. This property helps eliminate ambiguity when calculating embeddings based on anchor points.

To apply this method, the space is first divided into triangles. For example, the temperature-humidity space can be partitioned into triangles \( A_1A_2A_3 \) and \( A_2A_3A_4 \). Given a new sensor reading \( P \), the first step is to identify the simplex (triangle) that encloses \( P \). For instance, \( P \) might fall within the triangle formed by the anchors \( A_2 \), \( A_3 \), and \( A_4 \). The barycentric coordinates \((\alpha, \beta, \gamma)\) of \( P \) with respect to this triangle are then calculated. These coordinates quantify the contribution of each anchor to the interpolation.

Using the embeddings of the anchors, the embedding of \( P \), denoted as \( E_P \), can be computed as:
\begin{equation}
    E_P = \alpha E_{2} + \beta E_{3} + \gamma E_{4}.
\end{equation}

As the number of anchors increases, the sensor reading space can be subdivided into more simplices, enabling more refined and precise interpolation. For instance, in Figure~\ref{fig:temp_humidity_a5}, introducing an additional representative temperature-humidity sensor reading, \( A_5 \), allows \( P \)'s embedding to be calculated within the triangle \( A_1A_2A_5 \).

The subdivision of the space is achieved using Delaunay triangulation~\cite{lee1980twodelaunay}, which generates a network of triangles by maximizing the minimum angle of each triangle. This approach minimizes the occurrence of thin, elongated triangles, ensuring a smooth and consistent transition in the interpolated embeddings.


This approach ensures that \( E_P \) serves as a semantically consistent representation, accurately capturing the combined influence of temperature and humidity. Unlike direct linear interpolation, barycentric interpolation resolves the issue of non-uniqueness by utilizing the enclosing simplex.

\subsection{Generalization to Multiple Sensors}

\begin{figure}
        \centering
        \includegraphics[width=0.9\linewidth]{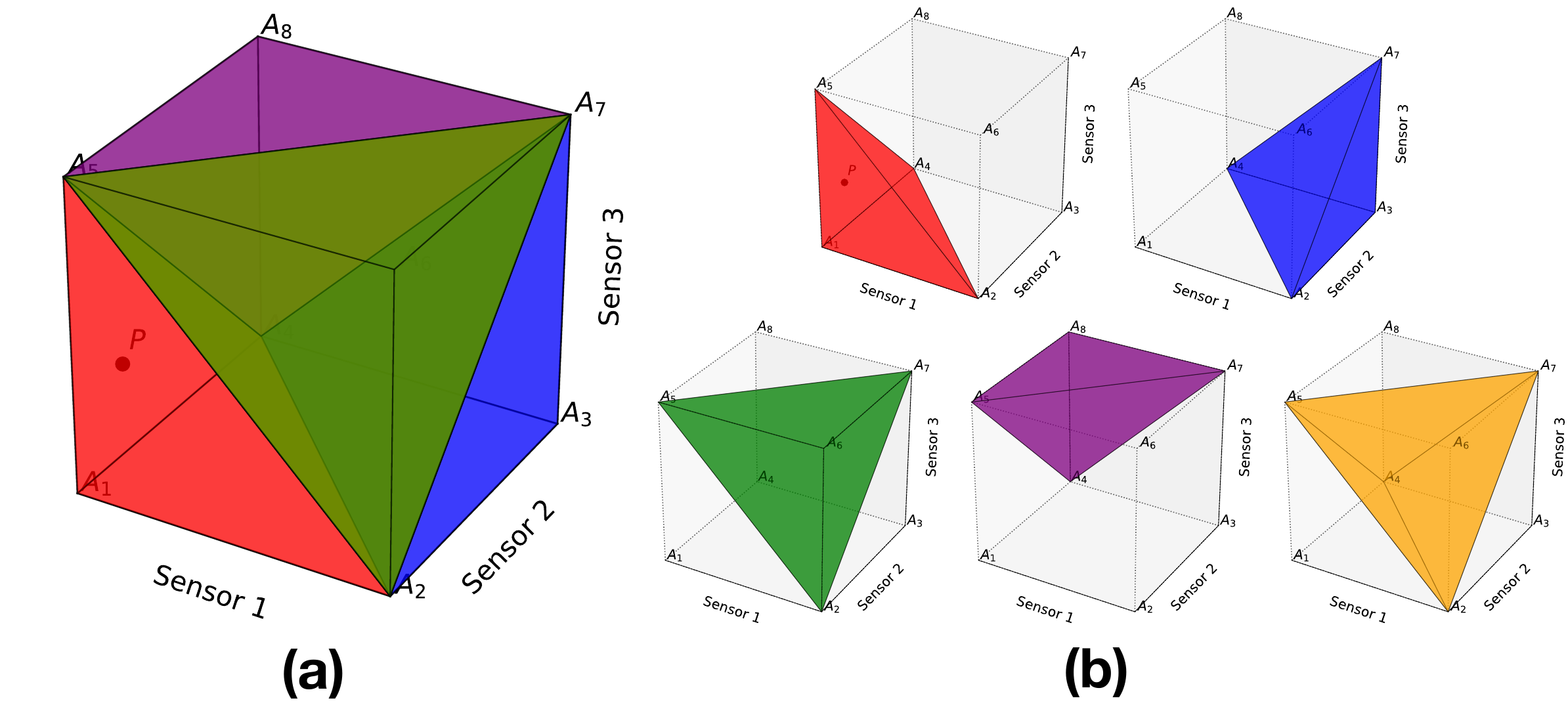}
        \caption{(a) The combination of readings from three sensors is divided into five tetrahedra for barycentric interpolation. (b) Show the divided tetrahedra.}
        \label{fig:temp_humidity_pressure}
\end{figure}

Our approach can be naturally extended to higher dimensions for scenarios involving more sensors. In such cases, the sensor space is divided into non-overlapping simplices (e.g., tetrahedra in 3D) using the generalized form of Delaunay triangulation, known as Delaunay tessellation~\cite{watson1981computing}, based on the anchor points.
Figure~\ref{fig:temp_humidity_pressure} illustrates that a space containing three sensors and eight anchors can be subdivided into five tetrahedra. The embedding \( E_p \) for a new sensor reading \( P \) can then be computed as a linear combination of the embeddings of the anchors \( A_1 \), \( A_2 \), \( A_4 \), and \( A_5 \).  
The generalized process for calculating embeddings is outlined below:

\begin{enumerate}
    \item \textbf{Tessellate the Sensor Space}: Partition the \(n\)-dimensional sensor space into non-overlapping simplices using Delaunay tessellation.
    \item \textbf{Identify the Containing Simplex}: Determine the simplex that encloses the new sensor reading point \(P\).
    \item \textbf{Calculate Barycentric Coordinates}: Represent \(P\) as a weighted combination of the simplex's vertices, \(P = \sum_{i=1}^{n+1} \alpha_i A_i\), where \(\sum_{i=1}^{n+1} \alpha_i = 1\) and \(\alpha_i \geq 0\).
    \item \textbf{Interpolate the Embedding}: Compute the embedding of \(P\) as \(E_P = \sum_{i=1}^{n+1} \alpha_i E_i\), where \(E_i\) are the embeddings of the vertices of the simplex.
\end{enumerate}

\paragraph{Achieved Features}
This approach ensures both coherence and scalability when embedding sensor data into a shared semantic space. The embeddings within each simplex are continuous, ensuring smooth transitions. Moreover, coherence is maintained across simplices, resulting in seamless and visually consistent embedding transitions as sensor values change, providing a coherent and smooth representation throughout the space. Additionally, the method scales efficiently with the number of sensors and anchors, enabling reliable interpolation across the different sensor modalities. 



\section{Sensor-Aware Visual Production}

\begin{figure}[t]
    \centering
    \includegraphics[width=0.9\linewidth]{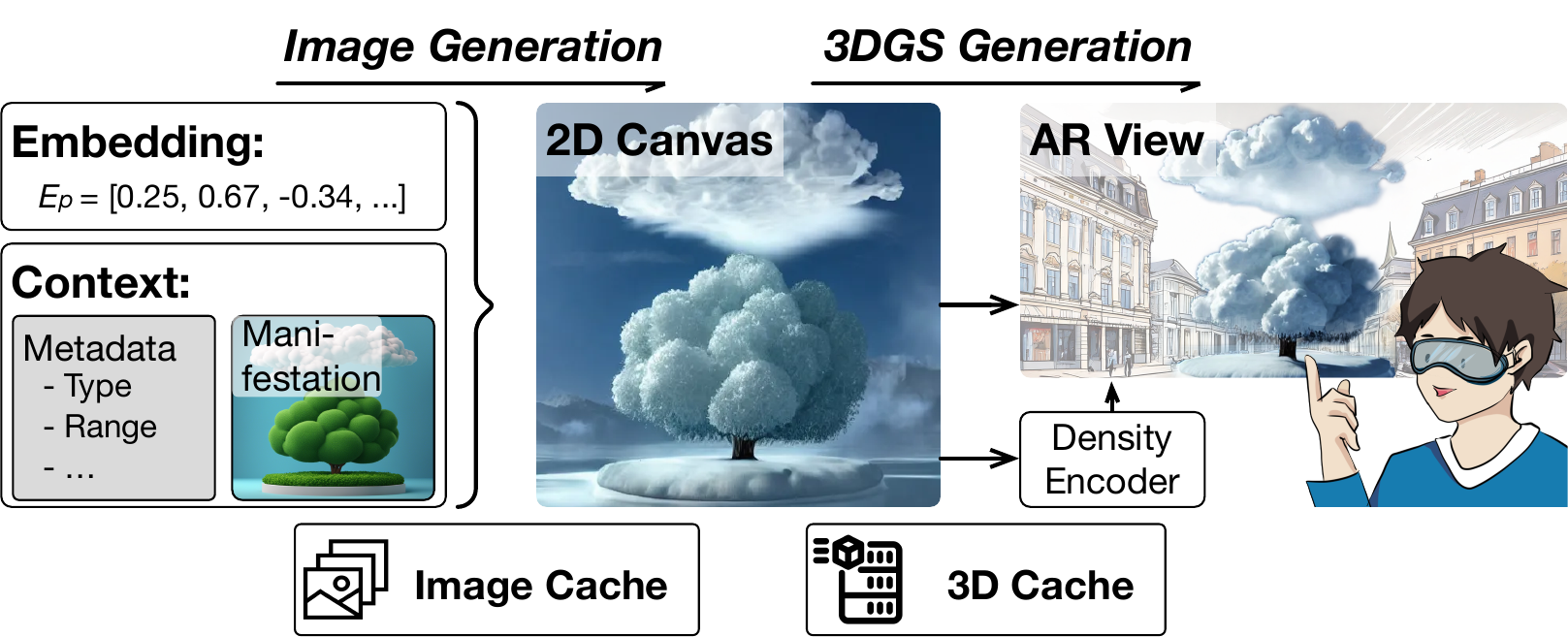}
    \caption{Sensor embedding to AR visualization.}
    \label{fig:visualization_workflow}
\end{figure}

Section~\ref{sec:design1} demonstrates how multi-modal sensor data can be embedded into a shared space, aligning sensor readings with visual embeddings through CLIP. Building on this foundation, this section introduces the sensor-aware visualization system, which explores methods to enrich data embeddings with sensor-specific features and leverages foundation models to transform these embeddings into coherent visual content.

Existing image generation models, such as Stable Diffusion, demonstrate the capability to generate visual content from CLIP embeddings~\cite{huggingface2023stablediffusion}, and these images can be further used to generate into 3D scenes using text-to-3D and image-to-3D models~\cite{tang2023dreamgaussian,poole2022dreamfusion,liu2023zero1to3}. However, sensor data is highly context-dependent and features volumetric and dynamic characteristics that current embedding-to-image and 3D generation models are not equipped to handle.

To overcome these challenges, we leverage the foundation models to incorporate domain knowledge and sensor schema to enrich sensor data to produce 2D visual representations. We further employ a customized 3D Gaussian Splatting (3DGS) model to encode spatial density, enabling the visualization of sensor data in 3D space. Additionally, to support low-latency visualization of dynamic sensor data, we design latent reuse caches at each step, enabling just-in-time AR presentations (Figure~\ref{fig:visualization_workflow}).

\subsection{Sensor Schema and Manifestation}
\label{sec-manifestation}

While embeddings from multi-modal sensor data coherently represent readings, they lack the contextual richness needed for intuitive visualizations. This can result in inconsistent and ambiguous visuals, undermining clarity and reliability. To address this, we propose a structured sensor schema and corresponding manifestations to ensure consistent, interpretable visualizations.

\paragraph{Sensor Schema}

A sensor schema structures data from diverse sources like home sensors, air quality metrics, weather data, and emotion detection. It consists of: (1) Sensor Types: A catalog of readings (temperature, humidity, emotional states); and (2) Context: Metadata about environment, purpose, subject, and units.

\paragraph{Manifestations}

Manifestations are real-world objects or symbols representing sensor data (e.g., clouds for weather, faces for emotions). The translation process involves: (1) Identifying Manifestations: Create consistent visuals using curated examples and LLMs; (2) Generating Constraint Maps: Create RGB and depth map templates; and (3) Visual Production: Use ControlNet~\cite{zhang2023adding}  with consistent seeds  to integrate sensor data into visuals that align with schema.

\begin{figure}
    \centering
    \includegraphics[width=0.9\linewidth]{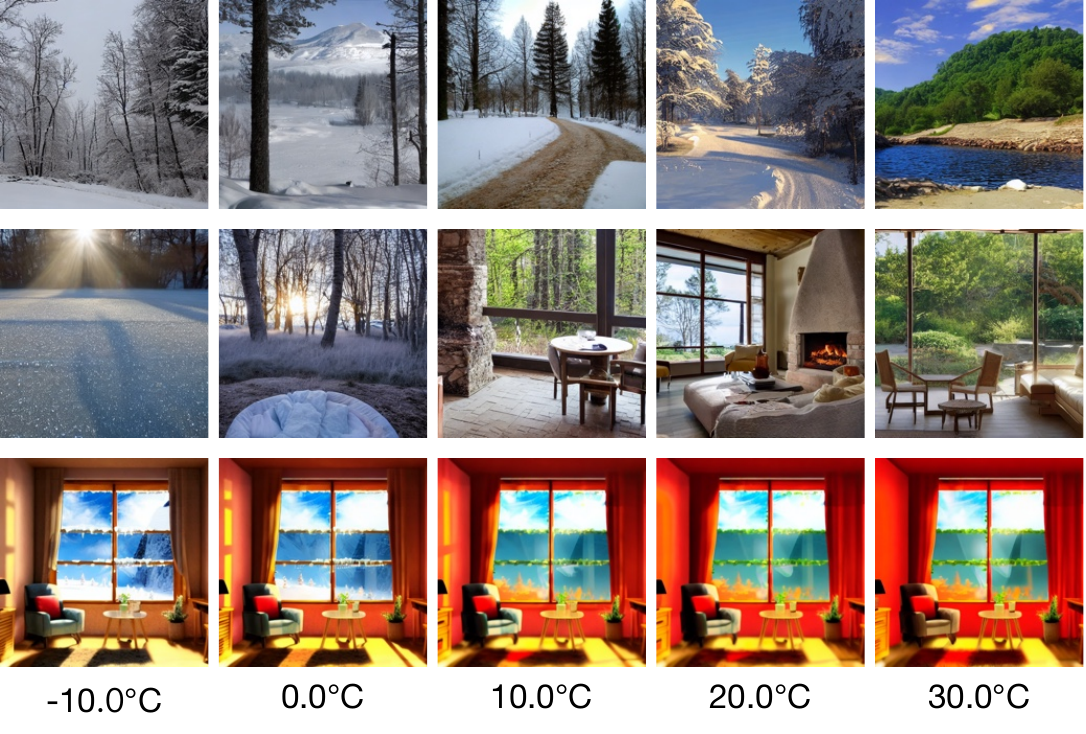}
    \caption{Comparison of our approach: the Stable Diffusion baseline (top), without manifestation (middle), and \pname 2D generation with `A Calm Room' manifestation (bottom).}
    \label{fig:showcase-room}
\end{figure}

Figure~\ref{fig:showcase-room} compares our method with baselines for temperature readings (-10$^\circ$C to 30$^\circ$C). Stable Diffusion~\cite{rombach2022high-stablediffusion} often misrepresents data (e.g., snow at 20$^\circ$C). Our embeddings improve accuracy but lack consistency, while the final visuals using the manifestation "A Calm Room" achieve both.

\subsection{AR Production}  


\paragraph{Volumetric Production with 3DGS}
From the CLIP embeddings and sensor context, we produce 2D visualizations. Building on this foundation, we extend these images into 3D AR content.
Traditional 3D content generation typically uses mesh-based models that represent only surface structures, leaving interiors empty. This approach is not suitable for sensor data, which often encapsulates volumetric phenomena. For example, environmental readings like temperature or humidity demand a representation that is volumetrically consistent rather than a hollow shell.  

To address this limitation, we develop a 3D generation approach that incorporates sensor embeddings and manifestations to produce volumetric models that accurately reflect the spatial context of sensor data.
Leveraging 3DGS~\cite{kerbl3Dgaussians, liu2023zero1to3, tang2023dreamgaussian}, we generate spatially-aware visualizations that capture the volumetric and spatial aspects of sensor data, such as temperature distributions within a space or air quality surrounding a street.

Our method integrates sensor embeddings and manifestations into the 3D generation pipeline by adapting the DreamGaussian~\cite{tang2023dreamgaussian} framework. We use sensor-specific embeddings and 2D images as inputs to generate volumetric representations that preserve the spatial characteristics of the sensor data. To create truly volumetric visualizations, we introduce a density-aware generation technique that applies controlled random dropout to Gaussian points, creating transparent areas that allow the model to fill both internal and external spaces. The dropout rate is determined by the sensor schema and readings through a density encoder, ensuring alignment with the sensor data's characteristics. This design ensures the generated AR content's interior is fully populated with Gaussian points.

\paragraph{LLM for Automated Production}
Embedding heterogeneous sensor data into visual spaces poses significant challenges, particularly in extreme conditions. Our proposed FM-assisted pipeline addresses these issues by leveraging large language models (LLMs) and schema generation. 
LLMs offer a robust solution for addressing the complexity and diversity of sensor data. These models excel in processing and understanding the intricate relationships between various data modalities. A key feature of our pipeline is the creation of a sensor schema that includes anchors specific to each sensor type and context. This schema is constructed using a combination of data collection, domain knowledge, and the capabilities of LLMs, enabling zero-shot interpretation by generating meaningful visual manifestations without prior task-specific training.

Our pipeline is specifically designed to handle sensor data using zero-shot learning capabilities, which allows for the automatic processing of new sensor inputs. We construct a schema for each sensor type that anchors the sensor data to its contextual meaning, leveraging LLMs for schema refinement. With careful prompt design, LLMs automatically generate sensor metadata descriptions, which are then visualized. Additionally, advanced image generation models are employed to create detailed and accurate visual constraints, ensuring consistency with the defined sensor schema and manifestation.


\subsection{Latent Reuse for Fast Generation}



The generation of both images and 3D Gaussian Splatting is computationally expensive. Even with high-end GPUs like the Nvidia RTX 4090, creating a single image can take over 5 seconds per frame, while generating a 3D scene with DreamGaussian~\cite{tang2023dreamgaussian} may require up to 50 seconds. These time demands significantly impact the practicality of sensor visualizations, especially when dealing with diverse and rapidly changing data. To mitigate this, we designed a \textbf{latent reuse caching mechanism} that accelerates the generation of both images and 3D scenes, achieving a $7\times$ reduction in processing time without compromising output quality.

Images generated using ControlNet from neighboring embeddings tend to exhibit substantial similarity. Instead of starting from a random latent state (typically initialized from a Gaussian distribution), reusing the latent state from a previously generated image significantly reduces the number of iterations required to produce a new image that represents the corresponding embedding.

To implement this, we created a \textbf{cache table} that stores generated images along with their corresponding sensor readings. When new sensor data arrives, the system searches for similar (neighboring) readings in the cache. During the generation process, the cached neighboring image is used as the initial latent state, and the required number of iterations is determined based on the distance between the new sensor data and the cached reading.

This approach reduces the iterations needed for image generation from 50 to as few as 2–10, resulting in up to a 25$\times$ improvement in processing speed for new sensor readings. This optimization dramatically enhances the responsiveness and usability of sensor visualization systems.

\section{Evaluation and User Study}
\label{sec:user-study}


We deployed our system on a commodity server and AR devices, conducting evaluations to assess system performance, generation quality, and user feedback. The user study provided insights into the overall experience and potential application scenarios.

\subsection{Implementation}
\label{sec:implementation}


We built an end-to-end system from sensor reading to 3D AR production on commodity AR headsets. A server collects data from home sensors (e.g., thermometers, humidity sensors) and smart cameras (e.g., facial emotion extraction), and also integrates remote and online resources, such as weather data and observatories. 

We use a web-based framework based on Three.js~\cite{threejs, splat3dgs} and WebXR~\cite{wikipedia_webxr}, ensuring cross-platform compatibility across various AR devices, including Web, Mobile AR, AR headsets. We customized the shaders to support dynamic 3D Gaussian Splatting updates. We use \textit{gradio.app} to host the interaction APIs.

\paragraph{Embedding and Foundation Models}
We implemented the sensor embedding and volumetric production pipeline on the server using Python. The tokenizer and encoder are based on OpenAI's CLIP with ViT-L/14 Transformer architecture~\cite{radford2021learning}. For manifestation and image generation, we utilized state-of-the-art open-source models: SDXL~\cite{podell2023sdxl}, Stable Diffusion 1.5~\cite{rombach2022high-stablediffusion}, and ControlNet~\cite{zhang2023adding}. Our language model interactions use Llama 3.1 (70B and 405B), with cloud services handling the large models due to local server limitations. 
We customized the embedding-to-3D module from DreamGaussian's implementation~\cite{tang2023dreamgaussian}.

\subsection{System Overhead}
For content generation, we evaluated our system on a high-performance server equipped with an AMD Ryzen 9 7950X3D 16-core processor and an Nvidia 4090 GPU, ensuring efficient processing across the entire pipeline. As shown in Table~\ref{tab:system_overhead}, initialization for each sensor combination runs only once. The latent reuse cache reduces generation time by \textbf{11$\times$}, significantly improving the system's efficiency and practicality for managing dynamic sensor data.
This enables our solution to handle dynamic data streams like air quality and home environment conditions (e.g., temperature, humidity, lighting) timely. This efficiency ensures a seamless and interactive user experience across various domains.

\begin{table}[t!]
\centering
\begin{tabular}{lcc}
\toprule
\textbf{Component} & \textbf{Baseline (s)} & \textbf{Reuse Cache (s)} \\
\midrule
Schema + Manifest. & 9.6  & 9.6  \\
Image Generation    & 4.0  & 0.3  \\
3D Generation       & 54.5 & 5.0  \\
\textbf{Total Generation} & \textbf{58.5} & \textbf{5.3} \\
\bottomrule
\end{tabular}
\caption{System overhead reduction with latent reuse.}
\label{tab:system_overhead}
\end{table}

To evaluate rendering overhead for visualization, we tested various AR headsets—including the Meta Quest and Apple Vision Pro—as well as mobile AR platforms. Comprehensive testing across diverse user interfaces and hardware configurations demonstrated the adaptability and robustness of our solution. By constraining the Gaussian paint number to 10,000, we maintained frame rates above 30 FPS on Meta Quest, Vision Pro, and mid-range smartphones like the iPhone 13 Pro, demonstrating \pname's compatibility with common consumer devices.


\subsection{Evaluations of Visualization Quality}

\begin{figure*}
    \centering
    \begin{minipage}[b]{0.23\textwidth}
        \centering
        \includegraphics[width=\textwidth]{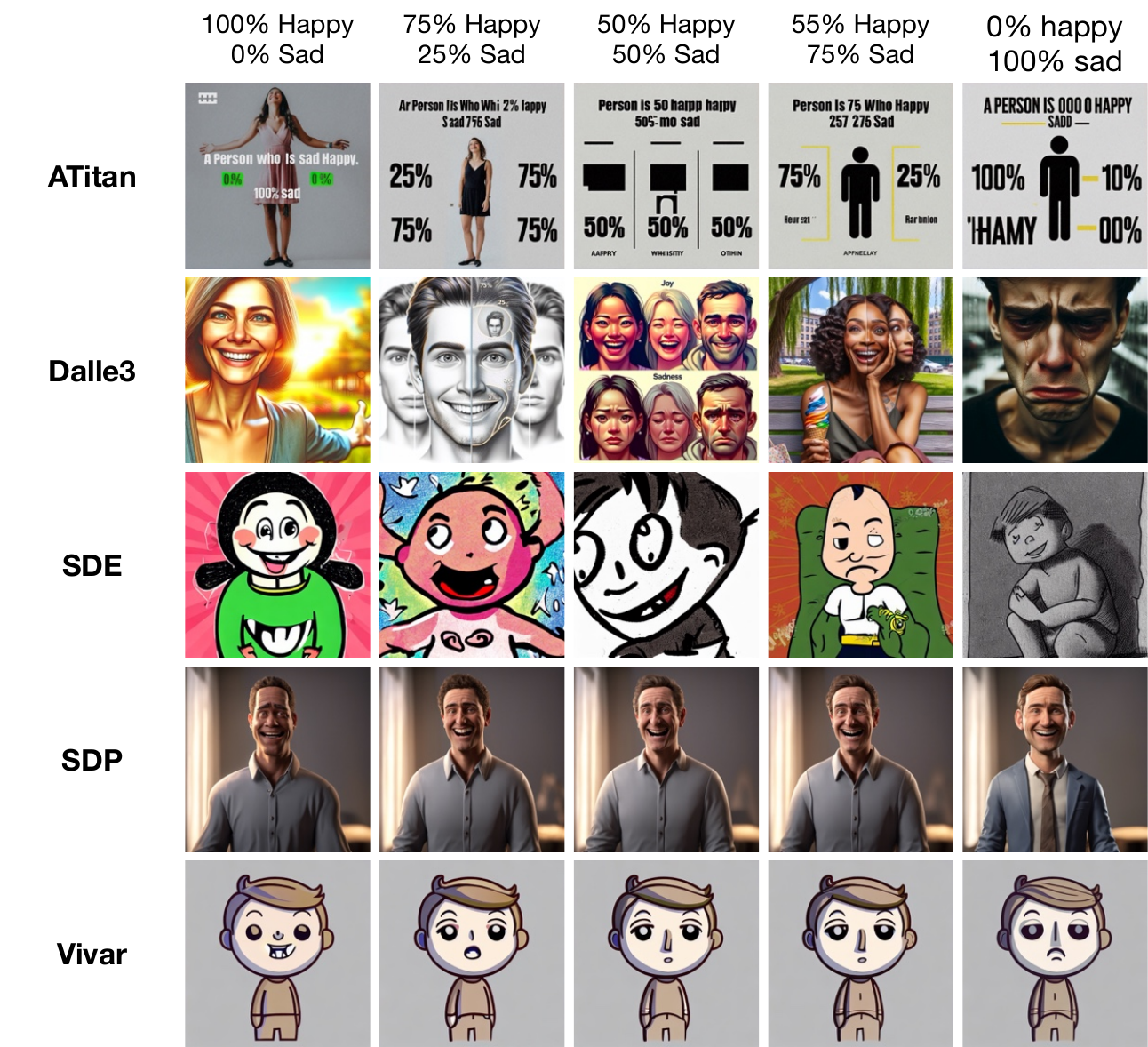}
        \caption{Visualization of human emotional states.}
        \label{fig:emotion-example}
    \end{minipage}
    \hfill
    \begin{minipage}[b]{0.33\textwidth}
        \centering
        \includegraphics[width=\textwidth]{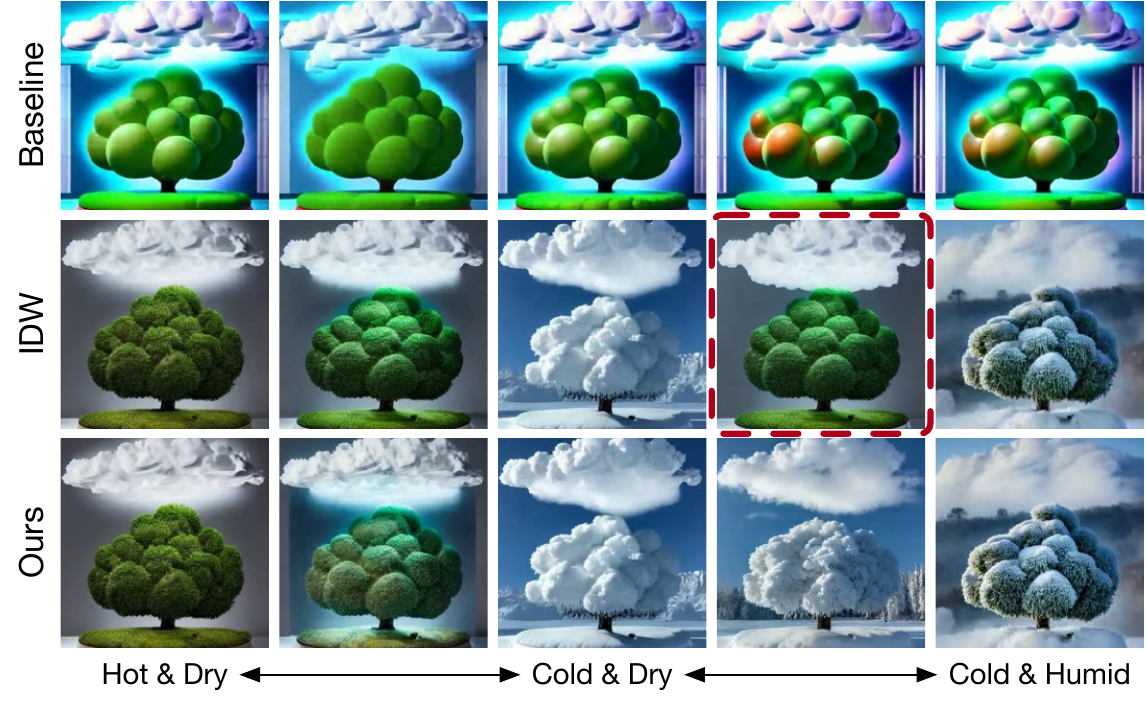}
        \caption{Environmental data visualization -- temperature and humidity.}
        \label{fig:weather-example}
    \end{minipage}
    \hfill
    \begin{minipage}[b]{0.40\textwidth}
        \centering
        \includegraphics[width=\textwidth]{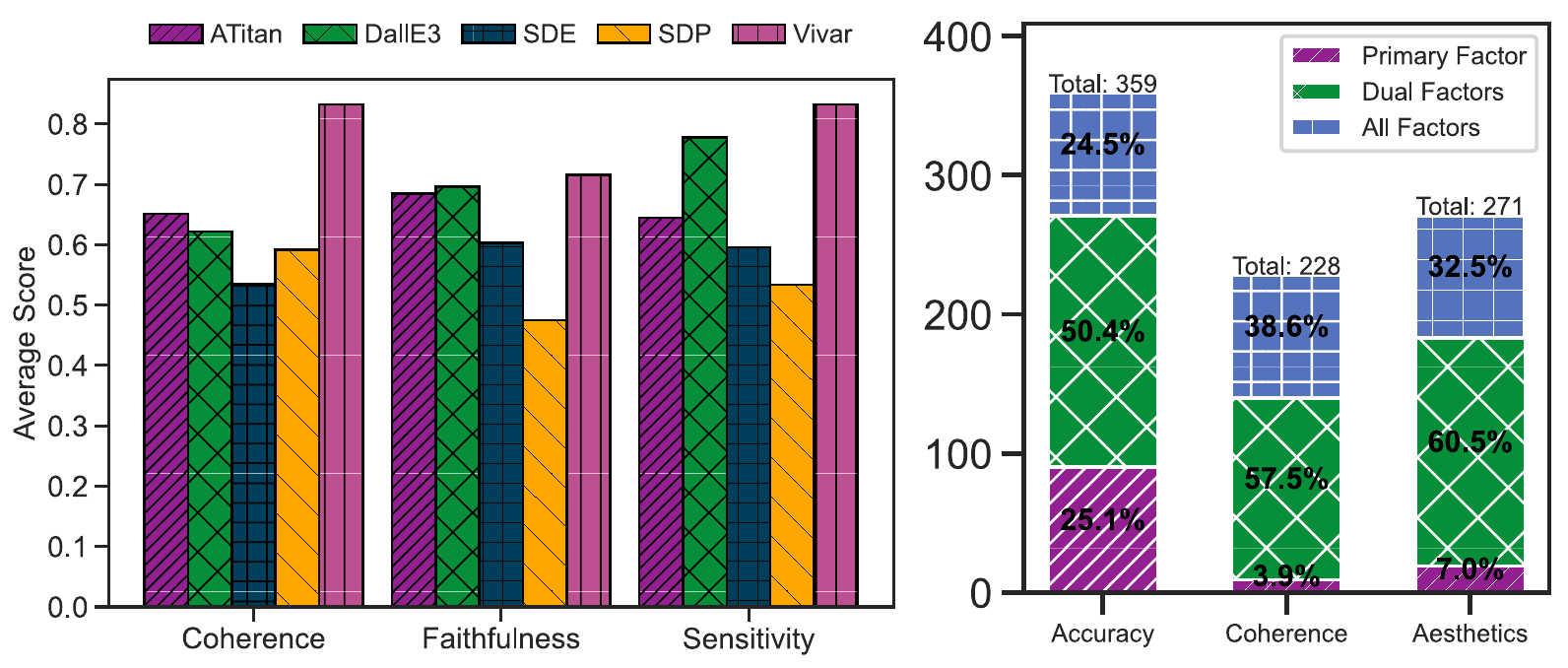}
        \caption{Model performance comparison across contexts (left) and user survey considerations (right).}
        \label{fig:model_performance}
    \end{minipage}
\end{figure*}

\paragraph{Experiment Design} To assess the quality of generated content, we conducted a survey with {565 participants} recruited through social media and forums, a total of 485 valid responses, of which \textbf{91.2\%} had prior experience with sensor technologies, enhancing their evaluation accuracy. The survey aimed to compare \pname\ with state-of-the-art image generation models based on three criteria, each rated on a scale from 1 to 5, where 1 is \textit{bad}, 3 is \textit{okay}, and 5 is \textit{awesome}: 
\begin{itemize} 
    \item \textbf{Sensitivity}: How well the images reflected variations in sensor readings. Higher scores indicate a more accurate reflection of changes in numerical values. 
    \item \textbf{Coherence}: The logical consistency between image transitions. 
    \item \textbf{Faithfulness}: How accurately the images represented the underlying sensor data. 
\end{itemize}


We compared our system, \pname, with two state-of-the-art generative models — Amazon Titan~\cite{aws_titan_models}
and DALL-E~3~\cite{betker2023improving} — as well as two Stable Diffusion models~\cite{rombach2022high-stablediffusion,podell2023sdxl} for \add{\textbf{ablation study}}. One Stable Diffusion model used blended embeddings as input (\textbf{SDE}), and the other relied on prompts with varying numerical values (\textbf{SDP})
To ensure consistency, we fixed the random seed during image generation and concealed the names of the methods from participants to avoid bias.




We evaluated the models across three contexts and for each context, participants were provided with five sets of five images, each set generated by one of the five models.:

\begin{enumerate} 
    \item \textbf{Environment}: Participants imagined being in a room with temperature and humidity sensors. The visualizations were intended to reflect these changes. 
    \item \textbf{Emotion}: Participants viewed images depicting a person experiencing a range of emotions (Figure~\ref{fig:emotion-example}). 
    \item \textbf{Entertainment}: Participants imagined listening to different types of music, from soft to noisy, expecting the visualizations to correspond to these auditory variations. 
\end{enumerate}

\begin{table}[t]
\centering
\resizebox{\columnwidth}{!}{%
\begin{tabular}{lcccc}
\toprule
\textbf{Model} & \textbf{Coherence} & \textbf{Faithfulness} & \textbf{Sensitivity} & \textbf{Overall Score} \\
\midrule
ATitan      & 0.65 & 0.68 & 0.64 & 0.66 \\
DallE3      & 0.62 & 0.70 & {0.78} & 0.70 \\
SDE         & \textcolor{Vermillion}{0.53} & {0.60} & 0.60 & {0.58} \\
SDP         & 0.59 & \textcolor{Vermillion}{0.48} & \textcolor{Vermillion}{0.53} & \textcolor{Vermillion}{0.53} \\
Vivar (Ours) & \textcolor{BluishGreen}{0.83} & \textcolor{BluishGreen}{0.72} & \textcolor{BluishGreen}{0.83} & \textcolor{BluishGreen}{\textbf{0.79}} \\
\midrule\midrule
Improvement & 0.18 (27.7\%) & 0.02 (2.9\%) & 0.05 (6.4\%) & 0.09 (12.9\%) \\
\bottomrule
\end{tabular}%
} 
\caption{Performance comparison across different evaluation metrics. The Improvement row shows the absolute and relative (\%) difference between \pname and the second-best model for each metric.}
\label{tab:survey_comparison}
\end{table}

\textbf{Overall Score:} To align our model evaluation with user priorities, we developed a weighted scoring system based on survey data. Participants indicated their most important factors when evaluating models, selecting from \textbf{Accuracy}, \textbf{Coherence}, and \textbf{Aesthetics}. For this analysis, we focused on \textbf{Accuracy} and \textbf{Coherence}, mapping Accuracy to \textbf{Faithfulness} and \textbf{Sensitivity}. Detailed breakdowns are provided in right figure of Figure~\ref{fig:model_performance}.

Weights were assigned based on the assumption that factors selected alone are more important than those selected with others. Specifically, weights were allocated as \( w_1 = 0.5 \) for factors selected alone, \( w_2 = 0.3 \) for factors selected with one other, and \( w_3 = 0.2 \) for factors selected with both others. The weight for each factor was calculated using:
\begin{equation}
    \omega_{\text{Factor}} = \frac{\sum_{k} w_k \times N_{\text{Factor}} \times p_{\text{Factor},k}}{\sum_{\text{All Factors}} \sum_{k} w_k \times N_{\text{Factor}} \times p_{\text{Factor},k}}
\end{equation}
where \( N_{\text{Factor}} \) is the total number of participants who selected the factor, \( p_{\text{Factor},k} \) is the proportion of participants selecting the factor under key \( k \), and \( w_k \) are the key weights. Based on our calculations, we obtained the weights: \( \omega_{\text{Coherence}} = 0.344 \), \( \omega_{\text{Faithfulness}} = 0.328 \), and \( \omega_{\text{Sensitivity}} = 0.328 \).
The overall score \( S_i \) for each model \( i \) was calculated using a weighted sum of the evaluation metrics:
\begin{equation}
    S_i = \omega_{\text{Faithfulness}} F_i + \omega_{\text{Sensitivity}} S_i + \omega_{\text{Coherence}} C_i
\end{equation}
where \( F_i \), \( S_i \), and \( C_i \) represent the Faithfulness, Sensitivity, and Coherence scores for model \( i \), respectively.

\paragraph{Results and Analysis} 

\add{Table~\ref{tab:survey_comparison} summarizes the processed scores for each model, with the highest scores for each metric highlighted in green and the lowest in red. To handle variation in rating preference, we applied response style adjustment through baseline normalization, subtracting the minimum score within each question-factor group per participant. This preprocessing preserved relative differences between ratings while establishing a common reference point. Our proposed system, \pname, achieved the highest scores across all metrics. The most notable improvement was in coherence (0.83, 0.18 higher than the second-best model), attributed to the continuity provided by embedding blending, which results in more logically consistent image generation. \pname\ also excelled in faithfulness (0.72), reflecting the effectiveness of our manifestation generation in accurately representing sensor data. For sensitivity, \pname\ (0.83) outperformed DALL-E~3 (0.78), despite the latter's strong language understanding capabilities, validating our design goal that visualizations accurately reflect changes in values.
Overall, \pname\ achieved the highest performance score (0.79), outperforming DALL-E~3 and ATitan by approximately 19.7\% and 12.9\% respectively, when accounting for user-prioritized criteria. 
}



\delete{\textbf{Ablation Study:} SDE and SDP are subcomponents of \pname. SDP uses conventional prompt-to-image generation, while SDE applies embedding blending without the manifestation component. As shown in Table~\ref{tab:survey_comparison}, SDP performed worst across all metrics. Introducing embedding blending (SDE) significantly improved faithfulness and sensitivity, demonstrating that embedding blending enhances the system's ability to generate more accurate and faithful visual representations compared to prompt-based approaches.}

\add{\paragraph{Multi-modality Handling} Figure~\ref{fig:weather-example} compares our barycentric interpolation approach with the baseline (using text prompts with Stable Diffusion) and inverse distance weight (IDW) interpolation. Our method correctly produces visualizations with appropriate semantics while maintaining consistency across the sensor space. When transitioning from cold \& dry to cold \& humid conditions, IDW interpolation fails to preserve the semantic information of cold weather, as it simply averages the embeddings and cannot preserve semantic correctness (circled in red). 
A user study involving an additional 18 participants who watched 386 frames of continuous visual productions showed unanimous agreement (100\%) that our approach provided superior coherence and faithfulness in visualization. All participants indicated a clear preference for our method over the baseline and IDW interpolation.
}



\subsection{User Study: Interact with \pname}
To further evaluate the usability and effectiveness of \pname, we allowed participants to interact with the system and provide detailed feedback. The survey focused on several key aspects: user satisfaction with the generated visual representations, the clarity of both 2D and 3D visualizations, the ease of use in manually adjusting the visual manifestations, and the overall effectiveness of the tool in visualizing complex sensor data. These insights provided a comprehensive understanding of the tool’s usability and performance.

%

\paragraph{Experiment Design}
Participants in this study were invited to use \pname with their own sensor data or a dataset relevant to their context. They had the option to select between automatic manifestation, where the system autonomously determines how to visually represent the sensor data, or manual manifestation, where they could customize the visual representation. After each experiment, participants completed an open-ended survey to share their experiences, including any challenges they encountered and their overall satisfaction. Due to the requirement of a headset for the 3D AR visualizations, only a limited number of users tested the 3D output, though we ensured consistency by also evaluating the 2D visualizations.


\begin{figure}
    \centering
    \includegraphics[width=0.99\linewidth]{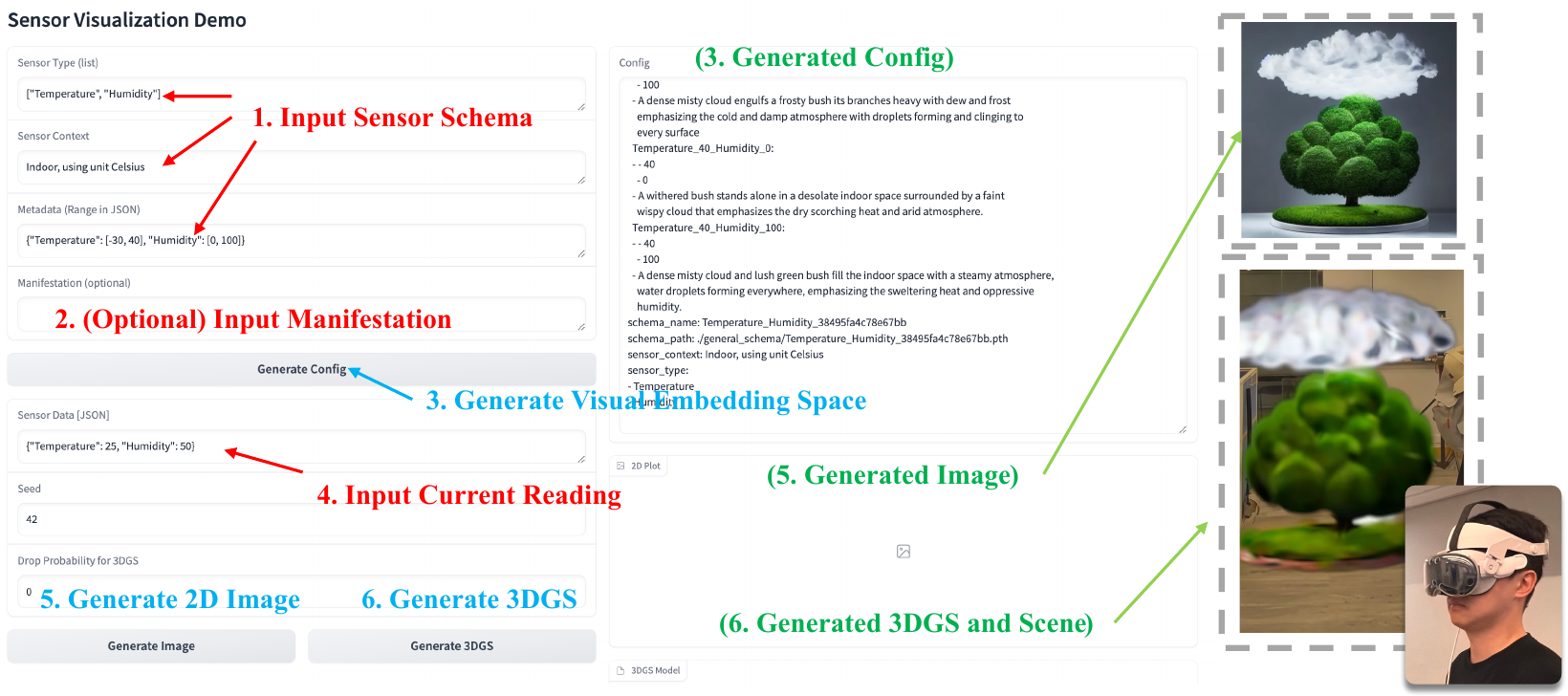}
    \caption{\pname user interface and AR setup.}
    \label{fig:gradio}
\end{figure}

\paragraph{Procedure and Tasks}
We recruited 37 participants to interact with \pname through a web-based platform. To attract participants, we posted QR codes at the entrances of lab buildings, enabling easy access to the system on personal devices. We developed a user interface (UI) that allows participants to input data and generate visualizations for their chosen parameters, as shown in Figure~\ref{fig:gradio}. Through this interface, participants can specify any type of sensor by providing the sensor type, context, and associated range of possible readings. Based on these inputs, the system generates a corresponding visualization embedding space. Optionally, participants can also provide a custom manifestation that serves as an input to the embedding space. If a manifestation is not specified, \pname will automatically generate one. After defining the parameters, the system then generates the correponsding 2D images and 3D objects when participants input sensor readings in JSON format.

After experimenting with \pname, participants were asked to answer the following questions:
\begin{itemize}
    \item \textbf{Q1:} Which type(s) of sensor information would you like to visualize through the system and how well? 
    \item \textbf{Q2:} How well does the manifestation produced by \pname represent the sensor data?
    \item \textbf{Q3:} When the sensor information changes, how easily can you recognize the difference in the generated visualization?
\end{itemize}


\paragraph{Result Analysis}
We recruited a total of 37 participants, allowing them to test \pname as many times as they wished. For each experiment, a cache was generated in the backend, and we collected a total of 190 records, excluding the default template provided for reference. On average, each participant interacted with \pname 5.1 times. In response to the first question (Q1), most participants initially tested ambient measurements, as these were more intuitive and familiar from daily life. As the study progressed, some participants began experimenting with more abstract visualizations, such as power consumption, precipitation, and motion data. 
Regarding the second question (Q2), more than half of the participants expressed satisfaction with the visual manifestations generated by \pname, describing them as vivid, intuitive, or contextually accurate. In response to the third question (Q3), nearly all participants reported that they could easily distinguish minor changes when adjusting the input values. However, a few participants noted that some cases, such as visualizing running vehicles, made it difficult to effectively reflect changes in speed using static images.

\paragraph{In-Depth Interviews with Industry Experts}
We conducted interviews with industry and educational domain experts to understand our system's strengths, limitations, and potential applications for addressing real-world challenges.

In the industrial context, we interviewed a senior hydrologist (P1) from a regional water management agency responsible for analyzing diverse hydrological sensor data, such as streamflow, groundwater levels, precipitation, and water quality.

P1 described the tools currently used for hydrological data analysis:
\begin{quote} 
    P1: \textit{"We rely on software like SWAT and HEC-HMS for hydrological modeling, MATLAB and Python libraries (e.g., pandas, SciPy) for statistical analysis, and GIS platforms like ArcGIS and QGIS for spatial visualizations. Temporal data is plotted using matplotlib and seaborn, while dashboards in Tableau or Power BI allow interactive stakeholder presentations."}
\end{quote}

This highlights the fragmentation in tools and the need for a unified system to integrate multiple data types. P1 emphasized the importance of continuous transitions and semantic consistency in visualizations:
\begin{quote} 
    P1: \textit{"Continuous transitions reveal variable evolution over time and space, essential for understanding dynamic processes. Semantic consistency preserves relationships between datasets, enabling accurate decision-making for tasks like flood prediction or pollution impact assessment. Errors or mismatches can lead to costly misinterpretations."}
\end{quote}

The interview underscores the critical need for integrated visualization tools that provide accurate, seamless, and unbiased representations, essential for effective hydrological analysis. Such tools could streamline workflows, improve decision-making, and support training by enhancing sensitivity to hydrological data while minimizing human error.

\vspace{-3mm}
\paragraph{In-Depth Interviews with Education Experts}

We interviewed two K-12 education experts, each with over a decade of experience in teaching and managing educational programs (P2 and P3), to explore how our system could enhance creativity and learning across age groups.

\paragrapht{Fostering Creativity in Younger Students}
P2 emphasized the system's potential to encourage creativity in early education:
\begin{quote}
P2: \textit{“This solution could foster creativity in younger students. For example, during model-building activities or competitions, it could simulate architecture or robotics. Younger children are especially focused on the creative experiences you mentioned.”}
\end{quote}
By enabling immersive simulations like architectural or robotic designs, our system aligns with experiential learning, helping young students engage in hands-on activities.

\paragrapht{Engaging Interactive Tools}
P2 also highlighted the system's potential as an interactive gadget:
\begin{quote}
P2: \textit{“It would be great if this could be a fun gadget, like inputting something to generate a realistic simulation—imagine placing an Ethiopian cloud on my desk. That kind of immersive experience would be amazing.”}
\end{quote}
This suggests the system could transform abstract inputs into tangible, playful simulations, creating memorable and engaging learning experiences for students.

\paragrapht{Advanced Learning for Older Students}
P3 discussed the value of the system for older students and professionals:
\begin{quote}
P3: \textit{“For older students or professionals in fields like biology, medicine, geology, or physics, this tool could be extremely beneficial. It needs to offer robust, discipline-specific functionalities for these advanced users.”}
\end{quote}

Her feedback highlights the system's potential to support specialized learning through accurate, consistent simulations tailored to advanced educational and professional needs.
The experts’ insights demonstrate the system's versatility: fostering creativity and engagement in younger learners while providing robust tools for specialized learning in older students and professionals, making it valuable across diverse contexts.

\section{Discussion}
\label{sec:discussion}

\paragraph{Downstream Applications}  
Our user studies highlight the versatility of our solution across diverse applications. In education, it translates abstract sensor data into immersive visualizations, making complex topics like climate patterns and chemical reactions more engaging and relatable, such as using environmental sensors to contextualize issues like air quality and climate change. In art and design, inspired by creators like \textit{Refik Anadol}~\cite{refik_anadol}, our method transforms sensor data into vivid visual expressions, enabling novel artistic works and interactive installations. In professional and industrial settings, it enhances data analysis and decision-making by visualizing sensor readings, such as predicting machinery maintenance in manufacturing. These examples demonstrate how our approach bridges complex sensor data and user-friendly visualizations, empowering diverse disciplines.

\paragraph{Future Work} 
Future exploration will focus on enhancing our system's capabilities to generate fully immersive and dynamic scenes, transcending the current object-based visualizations. This progression will necessitate more precise modeling of real-world environments and improved alignment between virtual objects and their physical counterparts. Additionally, integrating dynamic elements such as floating clouds and swaying trees following the physics can elevate the visual quality, creating a more lifelike and immersive experience that makes users feel truly present in the environment.


\section{Conclusion}
\label{sec:conclusion}

In this work, we addressed the challenge of visualizing multi-modal and context-dependent sensor data by introducing a novel cross-modal embedding method that blends sensor readings into a pre-trained visual embedding space, guided by the shading principle. By incorporating diverse manifestations and 3DGS, our approach enables accurate, consistent visual representations with smooth transitions between data points, maintaining semantic coherence. The inclusion of a latent reuse cache further enhances adaptability to dynamic sensor readings.
Through evaluations involving 503 participants, we demonstrated that our solution outperforms state-of-the-art visual generation models in consistency and representativeness, making it better suited to sensor data visualization. A user study with domain experts further validated the system's accuracy, reliability, and practical applicability across diverse downstream tasks. This work establishes a foundation for intuitive, coherent, and unbiased sensor data visualization by bridging domains. Drawing inspiration from graphics and shading principles, we combine large pre-trained models with explainable transformations, opening new avenues for advancing cross-modal embedding techniques.



\balance
\bibliographystyle{ACM-Reference-Format}
\bibliography{main}

\end{document}